\documentclass[twocolumn]{aastex631}
\usepackage{amssymb, amsmath}
\usepackage{graphicx}
\usepackage{natbib}
\usepackage{latexsym}
\usepackage{mathrsfs}
\usepackage{mathtools}
\usepackage{multirow}
\usepackage{xcolor}
\usepackage{color, colortbl}
\usepackage{booktabs,makecell}
\usepackage{float}
\usepackage{textcase}
\usepackage{censor}
\usepackage{array}

\newcommand{\tsup}{\textsuperscript}
\newcommand{\tsub}{\textsubscript}
\newcommand{\Lya}{\mbox{Ly$\alpha$}}
\newcommand{\K}{\mbox{K2-3}}
\newcommand{\E}{\texttt{EXOFASTv2}}

\usepackage{xcolor}
\definecolor{emphasis}{HTML}{bf2835}

\begin{document}

\title{The K2-3 system revisited: testing photoevaporation and core-powered mass loss with three small planets spanning the radius valley}

\correspondingauthor{Hannah Diamond-Lowe}
\email{hdiamondlowe@space.dtu.dk}

\author[0000-0001-8274-6639]{Hannah Diamond-Lowe}
\affiliation{National Space Institute, Technical University of Denmark, Elektrovej 328, 2800 Kgs.\ Lyngby, Denmark}

\author[0000-0003-0514-1147]{Laura Kreidberg}
\affiliation{Max Planck Institute for Astronomy, K\"onigstuhl 17, 69117 Heidelberg, Germany}

\author[0000-0003-2281-1990]{C.\ E.\ Harman}
\affiliation{Planetary Systems Branch, Space Science and Astrobiology Division, NASA Ames Research Center, Moffett Field, CA 94035, USA}

\author[0000-0002-1337-9051]{Eliza M.-R.\ Kempton}
\affiliation{Department of Astronomy, University of Maryland, College Park, MD 20742, USA}

\author[0000-0003-0638-3455]{Leslie A.\ Rogers}
\affiliation{Department of Astronomy, University of Chicago, 5640 South Ellis Ave., Chicago, IL 60637, USA}

\author[0000-0002-7403-7127]{Simon R.\ G.\ Joyce}
\affiliation{School of Physics and Astronomy, University of Leicester, University Road, Leicester LE1 7RH, UK}

\author[0000-0003-3773-5142]{Jason D.\ Eastman}
\affiliation{Center for Astrophysics $\vert$ Harvard \& Smithsonian, 60 Garden St., Cambridge, MA 02138, USA}

\author[0000-0002-3641-6636]{George W.\ King}
\affiliation{Department of Astronomy, University of Michigan, 1085 S.\ University Ave., 323 West Hall, Ann Arbor, MI 48109, USA} 

\author[0000-0002-5893-2471]{Ravi Kopparapu}
\affiliation{NASA Goddard Space Flight Center, 8800 Greenbelt Rd., Greenbelt, MD 20771, USA}

\author[0000-0002-1176-3391]{Allison Youngblood}
\affiliation{NASA Goddard Space Flight Center, 8800 Greenbelt Rd., Greenbelt, MD 20771, USA}

\author[0000-0002-6115-4359]{Molly R. Kosiarek}
\affiliation{Department of Astronomy and Astrophysics, University of California, Santa Cruz, CA 95064, USA}

\author[0000-0002-4881-3620]{John H.\ Livingston}
\affiliation{Astrobiology Center, 2-21-1 Osawa, Mitaka, Tokyo 181-8588, Japan}
\affiliation{National Astronomical Observatory of Japan, 2-21-1 Osawa, Mitaka, Tokyo 181-8588, Japan}
\affiliation{Department of Astronomy, University of Tokyo, 7-3-1 Hongo, Bunkyo, Tokyo 113-0033, Japan}
\affiliation{Department of Astronomy, The Graduate University for Advanced Studies (SOKENDAI), 2-21-1 Osawa, Mitaka, Tokyo, Japan}

\author[0000-0003-3702-0382]{Kevin K.\ Hardegree-Ullman}
\affiliation{Steward Observatory, University of Arizona, 933 North Cherry Ave., Tucson, AZ 85721, USA}

\author{Ian J.\ M.\ Crossfield}
\affiliation{Department of Physics and Astronomy, University of Kansas, Lawrence, KS, USA}

%max 250 words
\begin{abstract}
Multi-planet systems orbiting M dwarfs provide valuable tests of theories of small planet formation and evolution. K2-3 is an early M dwarf hosting three small exoplanets (1.5--2.0 R$_\oplus$) at distances of 0.07--0.20 AU. We measure the high-energy spectrum of K2-3 with HST/COS and XMM-Newton, and use empirically-driven estimates of Ly-$\alpha$ and extreme ultraviolet flux. We use \texttt{EXOFASTv2} to jointly fit radial velocity, transit, and SED data. This constrains the K2-3 planet radii to 4\% uncertainty and the masses of K2-3b and c to 13\% and 30\%, respectively; K2-3d is not detected in RV measurements. K2-3b and c are consistent with rocky cores surrounded by solar composition envelopes (mass fractions of $0.36^{+0.14}_{-0.11}\%$ and $0.07^{+0.09}_{-0.05}\%$), H$_2$O envelopes ($55^{+14}_{-12}\%$ and $16^{+17}_{-10}\%$), or a mixture of both. However, based on the high-energy output and estimated age of K2-3, it is unlikely that K2-3b and c retain solar composition atmospheres. We pass the planet parameters and high-energy stellar spectrum to atmospheric models. Dialing the high-energy spectrum up and down by a factor of 10 produces significant changes in trace molecule abundances, but not at a level detectable with transmission spectroscopy. Though the K2-3 planets span the small planet radius valley, the observed system architecture cannot be readily explained by photoevaporation or core-powered mass loss. We instead propose 1) the K2-3 planets are all volatile-rich, with K2-3d having a lower density than typical of super-Earths, and/or 2) the K2-3 planet architecture results from more stochastic processes such as planet formation, planet migration, and impact erosion.

\end{abstract}

\section{Introduction} \label{sec:intro}

Small exoplanets are currently divided into two broad categories: those smaller than about 1.7 R$_\oplus$ are dubbed super-Earths and are considered to be terrestrial in nature, with iron-silicate ratios similar to those of Earth and Venus, while those with radii between 1.7 and about 4 R$_\oplus$ have bulk densities consistent with degenerate amounts of iron, rock, ice and volatiles. The exoplanets in the second category are referred to as sub-Neptunes or gas dwarfs, given that their compositions must include some amount of low mean molecular weight material (hydrogen and helium) in order to explain their bulk densities. An extended hydrogen/helium atmosphere can make up only a few percent of a sub-Neptune's mass but contribute half of its observed radius \citep{Owen2020b}. Though we can measure the radii and masses of super-Earths and sub-Neptunes, these values can only provide a bulk density; we do not know the compositions of these worlds a priori. There is a well-measured valley in the radius distribution that separates the super-Earths and sub-Neptunes \citep{Fulton2017,Fulton2018,VanEylen2018,Martinez2019}, but the nature of the valley is not consistent across stellar types. 

For FGK stars there is a slope in the radius valley that trends towards larger planet radii with increasing instellation \citep[or decreasing orbital period;][]{Fulton2017,VanEylen2018,Martinez2019}. This change in location of the radius valley as a function of instellation gives rise to the narrative that small planets start out with hydrogen/helium envelopes but experience subsequent atmosphere sculpting. This can be explained by photoevaporation driven by the host star \citep{Owen&Wu2013,Lopez2013} and/or core-powered mass loss driven by energy left over from planetary formation \citep{Ginzburg2016,Ginzburg2018,Gupta&Schlichting2019}. 

On the other hand, low-mass stellar systems with $T_\mathrm{eff}<4700$ K observed by the Kepler and K2 missions show a tentative shallow trend in the radius valley towards larger planet radii with \textit{decreasing} instellation \citep[increasing orbital period;][]{Cloutier2020a}. This trend might be better explained by different formation timescales for sub-Neptunes and super-Earths: sub-Neptune worlds form rapidly ($<10$ Myr) and are able to accrete hydrogen and helium from the proto-planetary gas disc, while super-Earths take longer ($>10$ Myr) to form and do not have time to accrete gas from the disc before it dissipates \citep{Lopez2018}. However, gas remaining in the outer disc while the inner disc is depleted can seep into the inner disc while the terrestrial cores are forming (gas-poor formation), in which case the trend in the radius valley again resembles what is observed generally for FGK stars \citep{Lee2021}. It should be noted that with a temperature cut-off of 4700 K, the low-mass stellar sample includes both M and K dwarfs, and that the sample of planets is an order of magnitude smaller than that of FGK planets, so the planet radius valley for low-mass stars and its trend towards decreasing instellation is not definitive \citep{Cloutier2020a}.

Multi-planet systems allow for comparative planetology in the presence of a common star. The high-energy radiation from the stellar host can be factored out of the equation, so the different atmospheric outcomes of the planets are dependent only on their orbital distance. This is particularly powerful when we are unsure of the high-energy history of the stellar host, as is the case for most M dwarfs. M dwarfs are known to exhibit longer periods of activity in the pre-main sequence phase when compared to Sun-like stars and it is difficult to determine their ages, which would help to constrain the total amount of high-energy radiation an M dwarf planet experiences in its lifetime so far. In light of the difficulties presented by M dwarfs as small planet hosts, assessing the system architectures and planetary atmospheres of multi-planet M dwarf systems can provide evidence for the various theories of planet formation and evolution in the presence of these low-mass stars.

Here we take a deep dive into the \K\ system. \K\ is an early M dwarf ($M=0.55\mathrm{M}_{\odot}$, $T_\mathrm{eff}=3844\mathrm{K}$, $K_s=8.561$) hosting three small exoplanets \citep{Crossfield2015}. An early find by the K2 mission \citep{Howell2014}, this system is the subject of several follow-up observations to constrain the radii and masses of \K b, c, and d. Transit observations from K2 \citep{Crossfield2015} and Spitzer \citep[GO Program 11026, PI Werner; GO Program 12081, PI Benneke;][]{Kosiarek2019} constrain the radii of the \K\ worlds, while radial velocity measurements from ESO La Silla/HARPS \citep{Almenara2015}, Magellan II/PFS \citep{Dai2016}, TNG/HARPS-N \citep{Damasso2018}, and Keck I/HIRES \citep{Kosiarek2019} constrain their masses. \K b, the inner-most planet, has a radius and mass consistent with a sub-Neptune world. \K c, the middle planet, has a composition closer to terrestrial but has a slightly lower mass that points to the presence of extra volatiles. \K d, the outer-most world is not detected in the RVs but its radius of 1.5 R$_\oplus$ is consistent with what is expected for a terrestrial type world \citep[e.g.,][]{Rogers2015,Chen2017,Kanodia2019}. \K d also happens to orbit at the inner edge of the system's habitable zone \citep{Kopparapu2014}. 

In this work we present the high-energy spectrum of the host star \K\ and refine the planet parameters using a joint fit between transit, radial velocity, and SED measurements. We do not have measurements of the K2-3 planetary atmospheres so instead we test a range of atmospheres for each planet in the presence of the high-energy stellar spectrum in order to determine a possible range of atmospheric outcomes for these worlds. We include the effects of disequilibrium chemistry in these models, which is appropriate for worlds smaller and cooler than hot Jupiters, which can generally be assumed to be in thermal equilibrium.

In Section~\ref{sec:star} we describe the observations, analysis, and resulting data products of the high energy spectrum of the star \K. In Section~\ref{sec:exofast} we discuss the global fit that results in the planetary parameters used in this work. In Section~\ref{sec:models} we combine our knowledge of the stellar spectrum with interior and atmospheric models of the \K\ planets. We present a discussion of this work and its implications for the observed \K\ system architecture in Section~\ref{sec:discussion}. Concluding remarks can be found in Section~\ref{sec:conclusion}. 

The panchromatic spectrum of \K\ in various forms can be found as a MAST high level science product at doi: \dataset[10.17909/t9-fqky-7k61]{\doi{10.17909/t9-fqky-7k61}} (details in Section~\ref{sec:star}). 

\begin{deluxetable*}{cccccccc}
\centering
\caption{Observations of \K\ with HST/COS\label{tab:starobs}}
\tablewidth{0pt}
\tablehead{
\colhead{Visit} & \colhead{Orbit} & \colhead{Grating} & \colhead{Central Wavelength} & \colhead{Wavelength Range} & \colhead{Resolution Range\tsup{a}} & \colhead{Exposure Time} & \colhead{\texttt{FP-POS\tsup{b}}} \\
\colhead{} & \colhead{}      & \colhead{}             & \colhead{(\AA)}        & \colhead{(\AA)} & \colhead{($\times10^3$)} & \colhead{(s)} &
}
\startdata
  & 1          & G230L & 2950  &    1677--3215        &   2.1--3.9     & 4 $\times$ 60 & 1,2,3,4 \\
  \cline{2-8}
  & 1          &       &       &            &        & 706  & 3\\
  & 2          & G130M & 1291  &    1132--1430    &   12--16     & 2915 & 4 \\
\multirow{2}{*}{1} & 3          &       &       &            &        & 1682 & 3 \\
  \cline{2-8}
  & 3          &       &       &            &        & 996  & 1 \\
  & 4          & \multirow{2}{*}{G160M} & \multirow{2}{*}{1589}  &   \multirow{2}{*}{1395--1766} &  \multirow{2}{*}{13--20}      & 2915 & 2 \\
  & 5          &       &       &            &        & 2915 & 3 \\
  & 6          &       &       &            &        & 2915 & 4 
\enddata
\tablecomments{At the time these data were taken (14 November 2017) the G130M and G160M gratings were placed at lifetime position four (LP4) on the FUV detector to mitigate gain sag and extend the detector lifetime.\\
\tsup{a} Resolving power $R=\lambda/\mathrm{FWHM}$ from Table 5.1 of the COS Instrument Handbook v13\\
\tsup{b} Changing the fixed-pattern position (\texttt{FP-POS}) shifts the spectrum slightly in the dispersion direction such that spectral features fall on different detector pixels. This is done to overcome limitations in S/N achievable with each grating due to fixed-pattern noise in the COS detectors.
}
\end{deluxetable*}

\section{Panchromatic Spectrum of K2-3} \label{sec:star}

To construct the panchromatic spectrum of K2-3 we must combine information from multiple sources, including observations, estimates, and models. In this section we break down each piece of the spectrum and explain how we arrived at the flux density before putting it all together in a spectrum that spans 1--10\tsup{5} \AA.

\subsection{Ultraviolet} \label{subsec:uv}
We observed K2-3 in both the far-UV (FUV=912--1700\AA) and near-UV (NUV=1700-3200\AA) with the Cosmic Origins Spectrograph on board the Hubble Space Telescope (HST/COS). We observed for six orbits in a single visit on 14 November 2017 to obtain these data (GO Program 15110, PI L.\ Kreidberg). Details of the observations, including the grating, wavelength range, and exposure times can be found in Table~\ref{tab:starobs}. The HST/COS pipeline combines observations taken with the same grating and provides wavelength-calibrated 1D spectra for each grating. A resolution element of the COS FUV detector spans $6\times10$ pixels and for the COS NUV detector $3\times3$ pixels (COS Instrument Handbook v13 Chapter 3.2). We bin the 1D spectra in each grating by 4 pixels per bin for gratings G130M and G160M and by 3 pixels for G230L. We then combine the spectra from the three gratings and Doppler shift the spectrum in wavelength space according to the systemic radial velocity of \K\ \citep[30.76 km/s;][]{GaiaDR22018}. 

\subsubsection{Ly-\texorpdfstring{$\alpha$}{Lg}} \label{subsubsec:lya}

As part of the design philosophy for COS to achieve high throughput of point sources in the FUV, COS is effectively a slitless spectrograph with a large entrance aperture (2.5$''$ diameter) \citep{Green2012}. As a result, COS spectra suffer from significant geocoronal contamination originating in Earth's atmosphere at the N \textsc{i} (1200\AA)  \Lya\ (1214\AA), and O \textsc{i} (1302, 1305, 1306\AA) lines. We assume that measured N \textsc{i} and O \textsc{i} flux is entirely contaminated and we set the flux density in the narrow ranges around these lines (1198--1202\AA, 1301--1308\AA, and 1354--1357\AA) to 0 erg cm\tsup{-2} s\tsup{-1} \AA\tsup{-1}. The \Lya\ flux makes up 85\% of M dwarf FUV flux \citep{France2016} so it cannot be zeroed out without significantly affecting our understanding of the UV spectrum. In the case of some M dwarf observations in the UV it is possible to observe the red and blue wings of the \Lya\ line using HST/STIS, which has a narrower slit and therefore experiences less geocoronal contamination \citep{France2016}. However, the \Lya\ line core is still inaccessible due to absorption by neutral hydrogen in the interstellar medium. The \Lya\ line wing observations by HST/STIS allow for a reconstruction of the \Lya\ line \citep{Youngblood2016}, but such observations become prohibitively expensive for inactive mid-M dwarfs beyond $\sim$15 pc. For \K\ we instead use the UV lines that we do measure in the FUV and NUV along with known UV-UV flux correlations \citep{Youngblood2017} to estimate \K's \Lya\ line flux.

\begin{figure*}
\includegraphics[width=\textwidth]{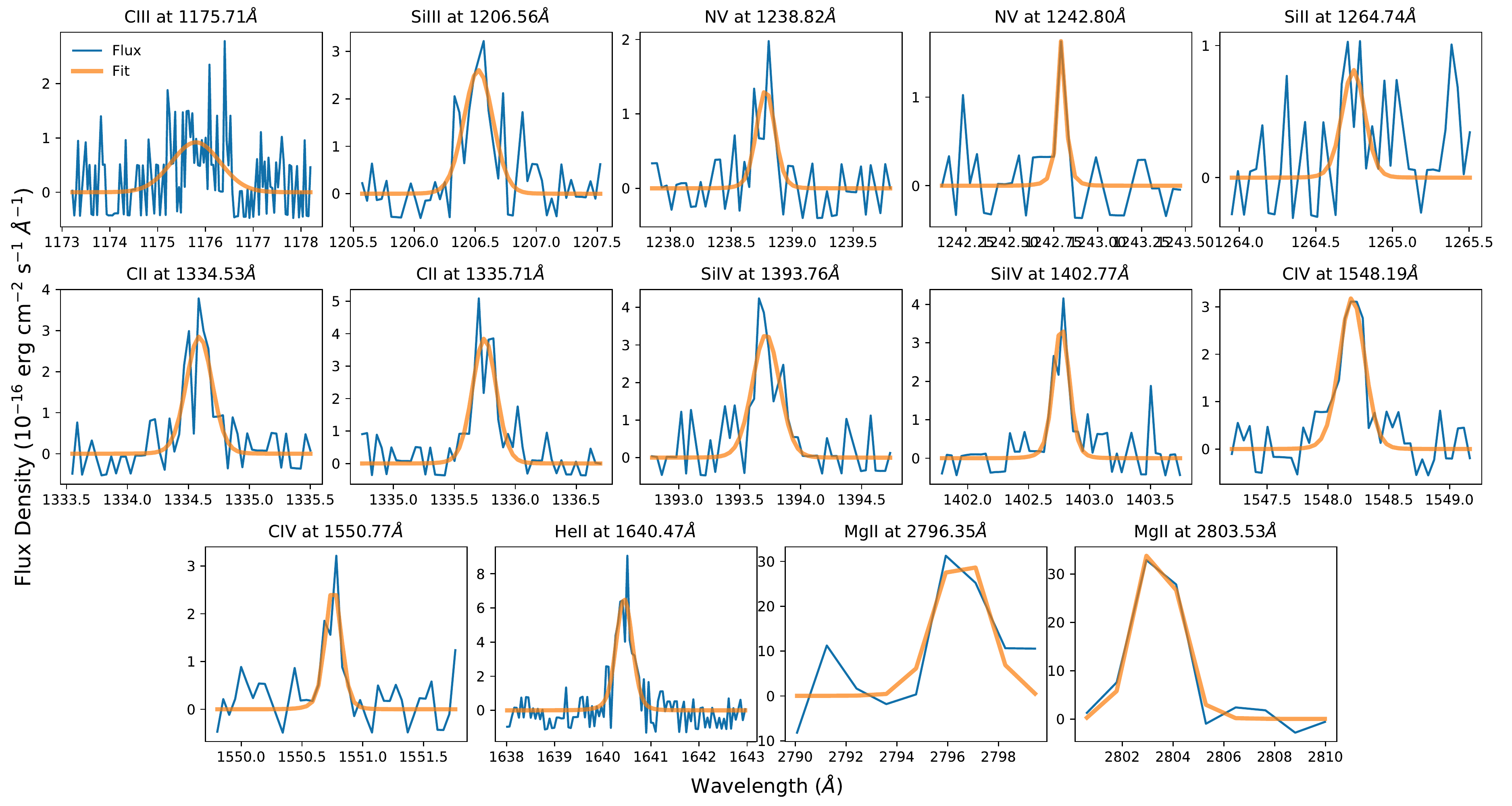}
\caption{Measured transition region lines (blue) listed in Table~\ref{tab:measuredlines}, with subplots for each transition. The transition lines are fitted with Gaussian functions convolved with COS LSFs (orange).}
\label{fig:measuredlines}
\end{figure*}

Similar to the procedure laid out in \citet{Diamond-Lowe2021}, we fit the observed UV lines with Gaussian profiles convolved with the appropriate COS line spread function (Figure~\ref{fig:measuredlines}) and then integrate under the resulting curve to determine the line flux. We scale these apparent fluxes by the distance and radius of \K\ to get their surface fluxes (Table~\ref{tab:measuredlines}). To estimate the surface flux of \Lya\ we use the measured surface fluxes along with the UV-UV correlation equations which take the form

\begin{equation}\label{eqn:UVscaling}
   \mathrm{log}_{10}(F_{\mathrm{Surf},\mathrm{ UV_1}}) = m \times \mathrm{log}_{10}(F_{\mathrm{Surf},\mathrm{ UV_2}}) + b
\end{equation}

\noindent where $F_{\mathrm{Surf},\mathrm{ UV_1}}$ is the \Lya\ surface flux we want to estimate, $F_{\mathrm{Surf},\mathrm{ UV_2}}$ is the surface flux of one of the measured transition region lines, $m$ is the fitted slope of the correlation between line 1 and line 2, and $b$ is the fitted $y$-intercept of the correlation. The variables $m$ and $b$ are provided in Table~9 of \citet{Youngblood2017} for the transition lines listed in Table~\ref{tab:measuredlines}. Our resulting estimates for \K's \Lya\ flux from each measured line are provided in Table~\ref{tab:Lyaestimates}. 

\begin{figure}
\includegraphics[width=0.48\textwidth]{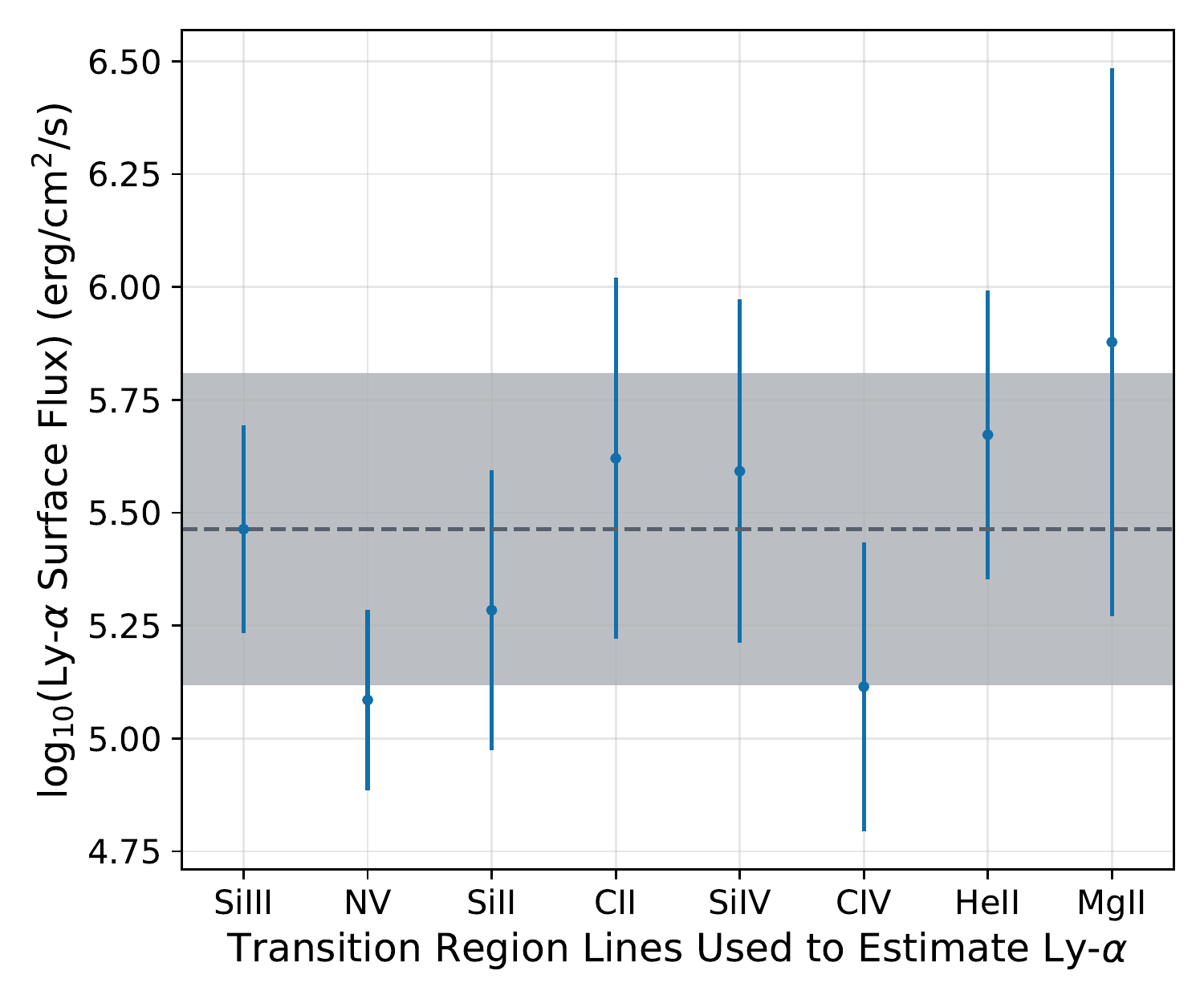}
\caption{\Lya\ estimates with 1$\sigma$ uncertainties (blue data points) calculated from each of the measured transition region lines in Table~\ref{tab:Lyaestimates} using Equation~\ref{eqn:UVscaling}. We take the weighted average as the \Lya\ flux and 1$\sigma$ uncertainty for \K\ (grey dashed line and shaded grey area).} 
\label{fig:lyaestimates}
\end{figure}

To determine the uncertainties in the \Lya\ estimate from each measured emission line we account for both the uncertainty in fitting profiles to the measured emission lines and the rms scatter in the UV-UV correlations. For each \Lya\ estimate from a measured emission line we compare the maximum (and minimum) \Lya\ values we would get from the propagating the flux density plus (or minus) its 1$\sigma$ uncertainty through equation~\ref{eqn:UVscaling}. We compare the result of the propagating the uncertainty to the rms scatter in the UV-UV correlation \citep[also provided in Table 9 of ][]{Youngblood2017}. Whichever value is bigger becomes the uncertainty in the \Lya\ estimate for that line. We then take the mean of the \Lya\ estimates and the mean of the \Lya\ estimate uncertainties as our final \Lya\ estimate and uncertainty. As can be seen in Figure~\ref{fig:lyaestimates}, the \Lya\ estimates from the measured emission lines agree well. 

We compare our \Lya\ estimate from the UV-UV correlations to a rough \Lya\ estimate from correlations with optical activity indicators \citep{Melbourne2020}. Starting from the H$\alpha$ equivalent width (EW) of $0.38\pm0.06$\AA\ reported by \citet{Crossfield2015}, we follow the procedures laid out in \citet{Melbourne2020} to translate this value into an estimate of the \Lya\ flux. We first determine the relative the H-$\alpha$ EW using Equations 2 and 3 of \citet{Newton2017}. We translate the relative H$\alpha$ EW into a luminosity relative to the bolometric luminosity $L_\mathrm{H\alpha}/L_\mathrm{bol}$ by multiplying by a factor $\chi$, which relates the continuum flux level for the H$\alpha$ line to the apparent bolometric flux of the star. We use a $\chi$ value of $6.533\times10^{-5}$ \citep[Table 8 of][]{Douglas2014}. We then calculate $\mathrm{log}_{10}(L_\mathrm{Ly\alpha}/L_\mathrm{bol})$ from $\mathrm{log}_{10}(L_\mathrm{H\alpha}/L_\mathrm{bol})$ according to the linear correlation provided in Table 4 of \citet{Melbourne2020}, and determine a value of $\mathrm{log}_{10}(L_\mathrm{Ly\alpha}/L_\mathrm{bol})=-4.67\pm0.32$, where we simply take the rms in the correlation as the uncertainty. Turning this value into a log\tsub{10}(\Lya\ surface flux) we get $5.68\pm0.32$, which agrees with our UV-UV correlation results (Table~\ref{tab:Lyaestimates} and Figure~\ref{fig:lyaestimates}).

When constructing the panchromatic spectrum we make a \Lya\ line profile based on the \Lya\ reconstruction model of \citet{Youngblood2016}, which is made from the combination of two Gaussian profiles. The first is a narrow Gaussian profile with FWHM = 138 km s\tsup{-1} and an amplitude of 90\% of the \Lya\ estimate, and the second is a broad Gaussian profile with FWHM = 413 km s\tsup{-1} and an amplitude of 10\% of the \Lya\ estimate. The sum of the two Gaussian profiles integrates to the estimated \Lya\ flux density. We splice this profile into \K's spectrum at the rest wavelength of \Lya\ (1215.67\AA) and spanning the wavelength range that shows contamination by geocoronal emission (1210--1221\AA).

\begin{deluxetable*}{c|ccccCC}
\centering
\caption{Measured Emission Lines from \K\ with HST/COS\label{tab:measuredlines}}
\tablewidth{0pt}
\tablehead{
\colhead{} & \colhead{Grating} & \colhead{Total Exposure Time} & \colhead{Line} & \colhead{Line Centers\tsup{a}} & \colhead{log$_{10}$(Surface Flux)} & \colhead{log$_{10}$($T$)\tsup{a}}\\
\colhead{} & \colhead{}      & \colhead{(s)}                 &                & \colhead{(\AA)}        & \colhead{(erg cm$^{-2}$ s$^{-1}$)} & 
}
\startdata
\multirow{9}{*}{\rotatebox[origin=c]{90}{FUV}} & \multirow{7}{*}{G130M} & \multirow{7}{*}{5303} & C \textsc{iii} & 1175.711\tsup{b} & $3.18\pm0.36$ & 4.8\\
                    &       &       & Si \textsc{iii} & 1206.555             & 3.01\pm0.16 &  4.7\\
                    &       &       & N \textsc{v}     & 1238.821, 1242.804  & 2.69\pm0.15 & 5.2\\
                    &       &       & Si \textsc{ii}   & 1264.738            & 2.30\pm0.36 & 4.5\\
                    &       &       & C \textsc{ii}    & 1334.532, 1335.707  & 3.35\pm0.15 & 4.5\\
                    &       &       & Si \textsc{iv}   & 1393.755, 1402.772  & 3.29\pm0.14 & 4.9\\
                    \cline{2-7}
                    & \multirow{2}{*}{G160M} & \multirow{2}{*}{9741} & C \textsc{iv} & 1548.187, 1550.775 & 3.22\pm0.14 & 5.0\\
                    &       &       & He \textsc{ii}   & 1640.474            & 3.57\pm0.06 & 4.9\\
\cline{1-7}
\multirow{2}{*}{\rotatebox[origin=c]{90}{NUV}} & \multirow{2}{*}{G230L} & \multirow{2}{*}{240}   & \multirow{2}{*}{Mg \textsc{ii}}   & \multirow{2}{*}{2796.352, 2803.531} & \multirow{2}{*}{5.43 ± 0.84} & \multirow{2}{*}{4.5}\\
&&&
\enddata
\tablecomments{Surface fluxes (erg cm$^{-2}$ s$^{-1}$) are calculated by scaling the observed flux by the distance $d$ and stellar radius $R_{\mathrm{s}}$ of \K: $F_{\mathrm{Surf}} = F_{\mathrm{Obs}} \times\ (d/R_{\mathrm{s}})^2 $. We use $d=44.13\pm0.11$ pc \citep{GaiaDR22018} and $R_{\mathrm{s}}=0.546\pm0.017\ \mathrm{R}_{\odot}$ (this work). For double lines, we report the combined surface flux. \\
 \tsup{a} Values from CHIANTI v7.0 \citep{Landi2012}\\
 \tsup{b} There are six unresolved C \textsc{iii} transition region lines here; we fit them as a single broad line
 }
 \end{deluxetable*}
 
\begin{deluxetable}{c|c}
\centering
\caption{\Lya\ Surface Flux Estimates from Measured UV Lines \label{tab:Lyaestimates}}
\tablewidth{0pt}
\tablehead{
\colhead{Emission} &  \colhead{log$_{10}$($F_{\mathrm{Surf}, \mathrm{Ly}\alpha}$)} \\
\colhead{Line} & \colhead{erg cm$^{-2}$ s$^{-1}$} 
}
\startdata
Si \textsc{iii} & $5.47\pm0.23$ \\
N \textsc{v}    & $5.09\pm0.20$ \\
Si \textsc{ii}  & $5.28\pm0.31$ \\
C \textsc{ii}   & $5.63\pm0.40$ \\
Si \textsc{iv}  & $5.59\pm0.38$ \\
C \textsc{iv}   & $5.11\pm0.32$ \\
He \textsc{ii}  & $5.67\pm0.32$ \\
Mg \textsc{ii}  & $5.88\pm0.61$ \\
\hline
Mean  & $5.46\pm0.35$ \\
\enddata
\tablecomments{C \textsc{iii} is not included in the UV--UV scaling relations \citep{Youngblood2017}.}
\end{deluxetable}

\subsubsection{Data gap and negative flux} \label{subsubsec:datagapbadflux}

There is a gap in the COS data from 2127--2767\AA\ due to the configuration of the G230L grating. In this case we also do not zero out the flux since it is apparent that there is a non-zero average flux on either side of the data gap, likely representing a small detection of the stellar continuum. We draw a sloped line from the average of the flux on one side of the gap to the average of the flux on the other side. Integrating across this slope will return an average flux value from both sides of the data gap. 

We find this an acceptable treatment of the data gap when compared to broadband UV coverage by the XMM-Newton and Swift observatories (further discussed in Section~\ref{subsec:xray}). The UV photometry from these observations spans parts of the data gap, and the measured fluxes are less than one order of magnitude greater than our estimate for the flux in the gap. The UV photometry should be considered an upper limit, though, since it includes prominent lines on either side of the data gap, including the He \textsc{ii} and Mg \textsc{ii} lines. Any lines in the data gap are relatively weak and would likely not have been visible above the COS noise floor for this target.

At many wavelengths the COS pipeline returns negative flux values, meaning that we are not able to detect any continuum flux from the star. However these negative values can be problematic for photochemical models that do not expect non-physical negative flux. Following the treatment of the MUSCLES spectra v2.2\footnote{\href{https://archive.stsci.edu/prepds/muscles/}{archive.stsci.edu/prepds/muscles/}}, we ``adaptively bin'' the data such that negative flux values are balanced by the positive flux in nearby wavelength bins \citep{France2016,Loyd2016}. This method conserves the overall flux across broad regions of the spectrum.

\subsection{X-ray} \label{subsec:xray}

\K\ was observed with XMM-Newton for a total 19 ks (Observation ID 0862060401, PI S.\ Joyce). The X-ray observations with the European Photon Imaging Camera pn CCD (EPIC-pn) resulted in a non-detection, but the star was clearly detected in the simultaneous UV observations with the XMM Optical Monitor. \textit{Swift} observations of K2-3 in June 2020 also resulted in a non-detection in the X-ray range, as would be expected given the lower sensitivity. From the XMM-Newton X-ray observations we place an upper limit on \K's X-ray flux. We calculate this upper limit by extracting a background subtracted count rate from the source region, using an extraction circle of 30$''$  radius which includes 90\% of the EPIC-pn point-spread function. The background count rate is measured from a larger region of 60$''$ to reduce error and is then scaled to match the size of the source region. Although no point source was detected, the count rate in the selected source region can still vary over time due to variations in the background flux. This was checked by plotting the lightcurve from both the source region and background measurement region. The pn exposure is 15.7 ks long and the region lightcurves are binned to 100 s time bins. 

It is clear from the background region lightcurve that the second half of the observation was affected by a systematically high background count rate. Because the background region lightcurve is subtracted from the source region lightcurve, the corrected source lightcurve shows high variability during the time when the background region count rate was high. A second issue is that the background subtraction appears to overcompensate, leading to some unphysical negative count rates in the corrected source region lightcurve. Because we do not make a detection of \K\ in the observations and are instead providing an upper limit on its X-ray emission, this overcompensation by the background subtraction would result in a lower value for the \K\ X-ray flux upper limit than we can actually determine from the observations. To mitigate this issue, we add 0.013 to the final background subtracted count rate in each bin to ensure that the resulting upper-limit is not unrealistically low. The corrected upper-limit on \K's X-ray flux is thus conservative.

The resulting count rate is then passed to PIMMs\footnote{\href{https://cxc.harvard.edu/toolkit/pimms.jsp}{cxc.harvard.edu/toolkit/pimms.jsp}} to calculate the flux. The model parameters used in PIMMs are:
\begin{itemize}
    \itemsep0em 
    \item Detector/grating/filter: XMM pn, none, medium
    \item Input energy: Set to default based on the energy range used for the XMM pn extraction
    \item Model: plasma/APEC
    \item Galactic neutral hydrogen column density: $5\times10^{18}$ cm$^{-2}$ (total column density); the column density per cm$^3$ is $\sim$0.1 cm$^{-2}$ \citep{Redfield2000}.
    \item Abundance: 1 (Solar)
    \item kT temperature: 6.1 (log($T$) Mega-Kelvin) / 0.1 (keV)
\end{itemize}

\noindent Since the plasma temperature parameter is unknown, we use a range of values for the Astrophysical Plasma Emission Code \citep[APEC;][]{Smith2001} temperature (0.1085--0.3058 KeV) appropriate for an inactive M-dwarf to produce a range of flux estimates. The upper end of the APEC temperature range provides the highest upper limit value, and we take this as \K's X-ray flux: $1.68\times10^{-14}$~erg~cm\tsup{-2}~s\tsup{-1} across the range 1.24--62\AA, using an upper limit count rate of 0.059 counts per second. When incorporating this value into the panchromatic spectrum we divide it across the wavelength range in order to convert the flux to a flux density in ~erg~cm\tsup{-2}~s\tsup{-1}~\AA\tsup{-1}. Integrating the panchromatic spectrum across the X-ray wavelength range returns the flux upper limit.

\subsection{Extreme ultraviolet} \label{subsec:EUV}

There are currently no operating observatories that can observe \K\ in the extreme ultraviolet (EUV=100--912\AA). We instead use a tool borrowed from the stellar astrophysics community \citep[e.g., ][]{Sanz-Forcada2003} that uses flux measured in the UV and X-ray to estimate the EUV. This approach is described in full detail in \citet{Duvvuri2021}. Briefly, it involves constructing the functional form of \K's differential emission measure (DEM), which describes the line-of-sight relative abundances of ions in the optically thin chromosphere, transition region, and corona under the assumption of collisional ionization equilibrium. 

\begin{figure*}
\includegraphics[width=\textwidth]{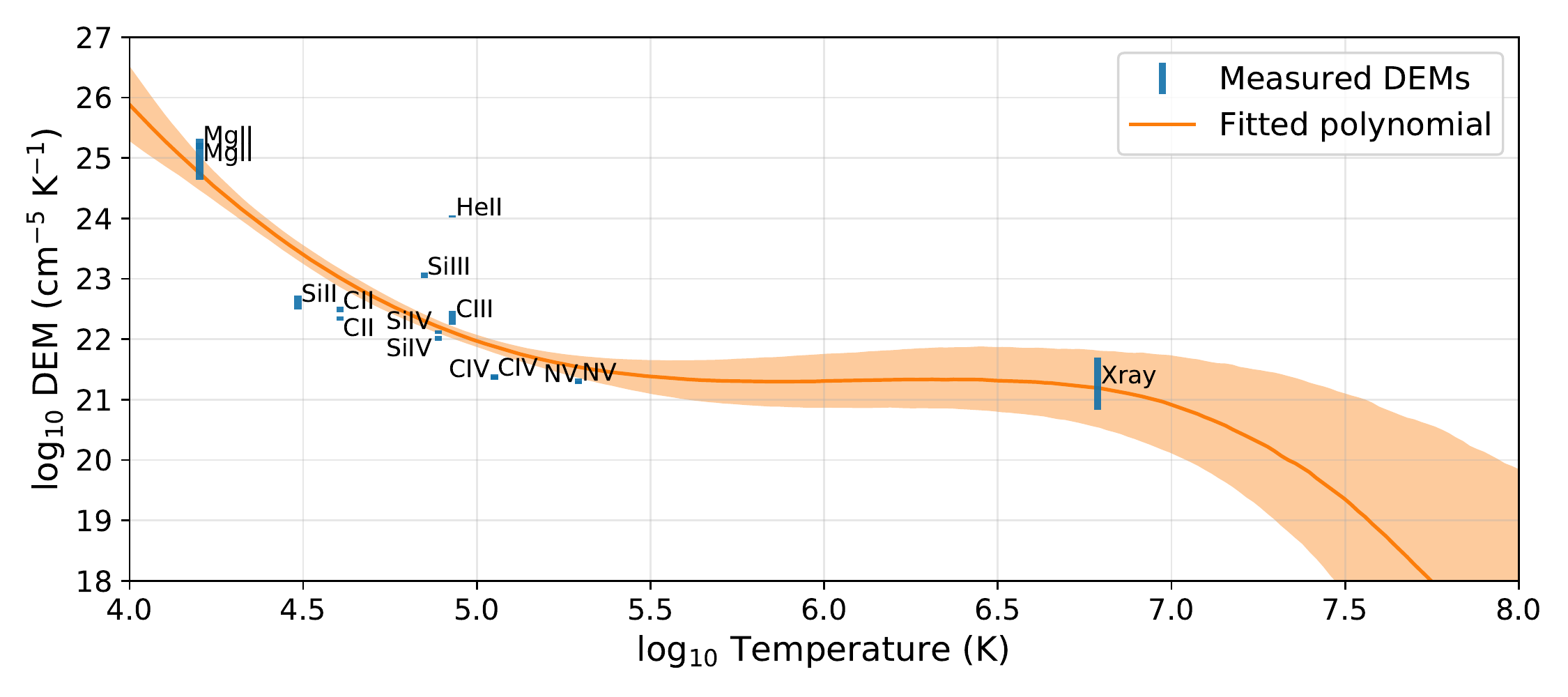}
\caption{Calculated differential emission measures with 1$\sigma$ uncertainties for observed transition region lines in the UV and broadband X-ray flux (blue bars). The best-fit DEM function is shown as an orange line with the shaded orange 1$\sigma$ region.} 
\label{fig:DEMs}
\end{figure*}

We calculate the DEM for each ion corresponding to measured transition region lines as well as the X-ray upper limit. To do this we calculate the intensity of each measurement and find the associated emissivity contribution function in the CHIANTI atomic database v10.0 using \texttt{ChiantiPy} \citep{Dere1997,DelZanna2021}. The X-ray upper limit corresponds to a broad wavelength range so in this case we combine the emissivity functions of all ions in that range. Following previous works \citep{Louden2017,Duvvuri2021} we then fit a 5\tsup{th}-order Chebyshev polynomial to the measured DEMs, using the open source dynamic nested sampler \texttt{dynesty} \citep{Speagle2020} to explore the parameter space. We use the same parameters and bounds as \citet{Duvvuri2021}. The best fit of that polynomial becomes the DEM function we use to estimate the unmeasured flux in the EUV. In Figure~\ref{fig:DEMs} we show the DEMs of the measured flux with uncertainties propagated from the line fitting described in Section~\ref{subsec:uv} at their peak formation temperatures, as well as the best fit polynomial function with 1$\sigma$ uncertainty.

When incorporating the EUV flux into the panchromatic spectrum of \K\ we divide the EUV region into broadband chunks of about 25\AA\ each from 62--912\AA, and 22\AA\ each from 912--1132\AA, after which point the HST/COS data come in. We make this divide at 912\AA\ because this is where the peak of \Lya\ continuum emission occurs, and is a natural place to divide the spectrum when integrating the EUV versus the FUV, for example. The different bin sizes are chosen to end in round numbers, which can help avoid small numerical differences when integrating different regions of the spectrum. Similar to what we do for the broadband X-ray data, we take the flux in each bin and divide it by the bin size in order to get the flux density for the panchromatic spectrum. 

\subsection{Optical and infrared}

To fill out the spectrum of \K, we append to the high-energy portion a BT-Settl model spanning the optical to mid-infrared \citep{Allard2014}, which extends to 10$\mu$m. From a grid of BT-Settl models we interpolate to the effective temperature and surface gravity of \K.

\subsection{Putting it all together}

We combine all of the pieces of \K's spectrum and provide four versions of the resulting panchromatic spectrum as high level science products, all available at doi \dataset[10.17909/t9-fqky-7k61]{\doi{10.17909/t9-fqky-7k61}}:
\begin{itemize}
    \itemsep0em 
    \item Variable resolution
    \item Constant resolution at 1\AA
    \item Variable resolution, adaptively binned to remove negative flux
    \item Constant resolution at 1\AA, adaptively binned to remove negative flux
\end{itemize}

\begin{figure*}
\includegraphics[width=\textwidth]{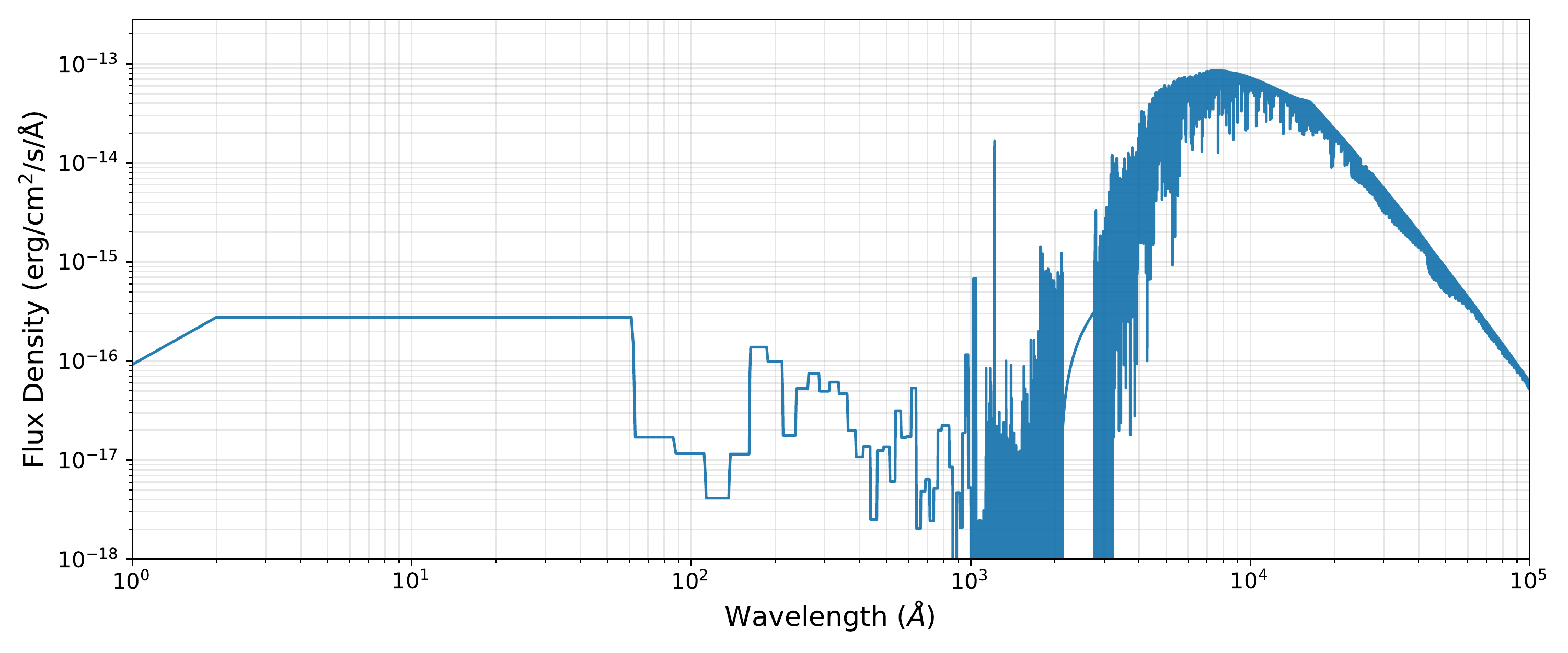}
\caption{Panchromatic spectrum of \K\ from 1--10\tsup{5}\AA\ and binned to 1\AA\ bins. A high level science product containing this spectrum can be found at doi: \dataset[10.17909/t9-fqky-7k61]{\doi{10.17909/t9-fqky-7k61}}}.
\label{fig:panchromaticspec}
\end{figure*}

\noindent These data products are designed to match those of the MUSCLES\footnote{\href{https://archive.stsci.edu/prepds/muscles/}{archive.stsci.edu/prepds/muscles/}} program \citep{France2016}; studies using the MUSCLES spectra should therefore be able to easily incorporate the spectrum of K2-3. We also provide the data from the COS G130M, G160M, and G230L gratings after we have binned the output data from the COS pipeline. The spectrum at constant resolution is shown in Figure~\ref{fig:panchromaticspec}.

\subsection{Comparison to other M stars}

We provide some benchmark properties of \K's high-energy spectrum in Table~\ref{tab:highenergyvalues}. \K\ fits with the high-energy trends for M dwarfs established by the MUSCLES survey \citep{France2016}, though no star in the MUSCLES sample is directly analogous to \K, an inactive early M star. For instance, \K\ falls closely in line with the trend in FUV/NUV flux ratio that decreases towards higher effective temperatures. With an effective temperature of $3844^{+61}_{-63}$ K, \K\ fills in a gap in the MUSCLES sample between early- to mid-M dwarfs and late K dwarfs \citep[see Figure 7 of ][]{France2016}. Finally, we calculate the energy-limited hydrodynamic mass loss rate for each of the \K\ planets as 

\begin{equation}
    \dot M = \frac{3\beta^2\eta F_{\mathrm{XUV,p}}}{4KG\rho_{\mathrm{p}}}
\end{equation}

\noindent where $\eta=\eta_\mathrm{eva}$, $\beta$ relates the planet radius to the radius at which XUV flux (= 1--912\AA) can be absorbed, $F_{\mathrm{XUV,p}}$ is the XUV flux at the planet, $K$ is a correction factor that accounts for mass only needing to reach the Hill radius to escape, and $\rho_\mathrm{p}$ is the planet density \citep{Erkaev2007,Salz2016}. In this calculation we use the XUV flux each planet currently receives from the host star \citep{Erkaev2007}, and we use prescriptions for the atmospheric expansion $\beta$ and escape efficiency $\eta_\mathrm{eva}$ from \citet{Salz2016}. We provide these values for each of the \K\ planet in Table~\ref{tab:highenergyvalues}. They will come into play later when we discuss the range of atmospheric cases we test for each planet in Section~\ref{subsec:atmoscases}.

\begin{deluxetable}{l|cccc}
\centering
\caption{Derived Values from the Spectrum of \K\label{tab:highenergyvalues}}
\tablewidth{0pt}
\tablehead{
\colhead{} & \colhead{Wavelength (\AA)} & \multicolumn{3}{c}{Value} 
}
\startdata
$\mathrm{log_{10}}(L_{\mathrm{Bol}})$ & 1--10\tsup{5} & \multicolumn{3}{c}{$32.35$} \\
$f$(XUV) & 1--912     & \multicolumn{3}{c}{$-4.39$} \\
$f$(FUV) & 912--1700  & \multicolumn{3}{c}{$-4.35$} \\
$f$(NUV) & 1700--3200 & \multicolumn{3}{c}{$-3.33$} \\
FUV/NUV & ---         & \multicolumn{3}{c}{$0.10$}  \\
\Lya/FUV & ---        & \multicolumn{3}{c}{$0.53\pm0.42$} \\ 
\hline
                            &        &  b                & c                & d \\
F$_{\mathrm{XUV}}$\tsup{a}  & 1--912 & $5.77\times10^2$  & $1.75\times10^2$ & $0.79\times10^2$ \\
$\beta$\tsup{b}             & ---    & 1.67              & 1.75             & 1.76 \\
$\eta_\mathrm{eva}$\tsup{b} & ---    & 0.26              & 0.31             & 0.32   \\
$\dot{M}$\tsup{c}           & ---    & $16.56\times10^8$ & $5.20\times10^8$ & $2.36\times10^8$
%$\tau$\tsup{d}             & ---    &            &            &   
\enddata
\tablecomments{Following \citet{France2016}, $f\mathrm{(band)}= \mathrm{log}_{10}(
L_{\mathrm{band}}/L_{\mathrm{Bol}})$; $F_{\mathrm{Bol}} = 9.68\times10^{-10}$ in all cases.\\
\tsup{a} The XUV flux (erg cm$^{-2}$ s$^{-1}$) at \K b, c, d is calculated using median values from a global computed with \E\ (Section~\ref{sec:exofast}, Table~\ref{tab:exofastvalues}). \\
\tsup{b} Values for atmospheric expansion $\beta$ and escape efficiency $\eta_\mathrm{eva}$ calculated according to \citet{Salz2016}. \\
\tsup{c} The hydrodynamic mass loss rate (g s$^{-1}$), also computed with median values from the global \E\ fit. In the case of \K d, which does not have a measured mass, we use the 3$\sigma$ observational upper limit of $2.2 \mathrm{M}_\oplus$. A higher mass for \K d would increase its surface gravity and decrease its hydrodynamic mass loss rate, thereby lengthening the timescale for mass loss. 
}
%\tsup{d} Time (Myr) to lose a hydrogen atmosphere weighing 1\% the planet mass.
\end{deluxetable}

\section{\K\ System Parameters} \label{sec:exofast}

\K\ was observed during the K2 mission, which took place after the initial Kepler mission ended due to failure of two out of four reaction wheels \citep{Howell2014}. The K2 phase of the Kepler mission did not point at the same patch of sky that Kepler stared at for four years, but rather moved around the ecliptic in multiple campaigns lasting roughly 80 days each. \K\ was identified in K2 Campaign 1 \citep{Crossfield2015}. 
The K2-3 planets have been the subject of numerous follow-up observations \citep{Kosiarek2019}. Because our goal is to model some possible atmospheres around these worlds we wish to measure the parameters of the \K\ system, and in particular the radii and masses of the planets, to as high a precision as possible given all of the available data. We update the \K\ system parameters using \E\ \citep{Eastman2019}. All of the data that go into this fit, along with the original publications, can be found in Table~\ref{tab:exofastdata}. We process the K2 light curves using \texttt{keplersplinev2}\footnote{\href{https://github.com/avanderburg/keplersplinev2}{github.com/avanderburg/keplersplinev2}} in order to remove instrument systematics and outliers. We mask out the transits when fitting the variation in the K2 light curve. Spitzer light curves and RV data are provided by \citet{Kosiarek2019}.

Much of the data in Table~\ref{tab:exofastdata} has already been published \citep{Kosiarek2019}, however the \E\ fit we present here incorporates two important updates: 1) we simultaneously fit the RV, transit, and stellar SED data in order to achieve our best-fit values, and 2) we include the updated distance from the Gaia Mission \citep{GaiaMission2016,GaiaDR22018}\footnote{The global \E\ fit was run before the most recent Gaia DR3 release. As there is little difference in the \K\ parallax between the Gaia DR2 and Gaia DR3 releases, and because we use the distance as a prior, not a fixed value, we did not re-run the global fit after the Gaia DR3 release.}. Fitting the planetary and stellar parameters simultaneously is important because these values are related to each other. For instance, the planet transits help to constrain their orbital inclinations, which in turn constrain the planetary masses from the radial velocity data. Transit and radial velocity information works together to constrain planet eccentricities. Including the high precision Gaia DR2 parallaxes to constrain the distance to \K\ affects the stellar luminosity and therefore significantly decreases the uncertainty in the stellar radius, since the luminosity is related to both the stellar distance and radius and is constrained by fitting the star's SED. Since transiting planet radii are measured in relation to the host star radius, decreasing the uncertainty in the stellar radius decreases the uncertainty in the planet radius. We also do not assume a zero eccentricity for the planetary orbits. Results of the \E\ fit can be found in Table~\ref{tab:exofastvalues}.

The \E\ code is a powerful and highly flexible tool for fitting exoplanet data sets. Based on the data provided and an initial set of priors, \E\ returns a set of posterior distributions for stellar and planetary values as well as jitter and offset terms that allow for simultaneous fitting over multiple data sets from multiple instruments. \E\ has been widely used and described in other work \citep[][ and references therein]{Eastman2019}, so we just highlight some of the specific settings we use:
\begin{itemize}
    \itemsep0em 
    \item We use the \texttt{/noclaret} flag in the \texttt{.pro} file (this is the main run script). The \texttt{/noclaret} flag turns off the limb-darkening interpolation, which is incomplete for M dwarfs. Without the interpolation we start from an initial guess at the limb-darkening parameters\footnote{\href{https://astroutils.astronomy.osu.edu/exofast/limbdark.shtml}{astroutils.astronomy.ohio-state.edu/exofast/limbdark.shtml}} \citep{Claret2011,Eastman2013} and add a wide uniform prior of $\pm0.15$.
    \item \E\ can use a stellar evolution model to inform the fit to the stellar parameters. In our fit we use the PAdova and TRieste Stellar Evolution Code \citep[PARSEC;][]{Bressan2012} by setting the \texttt{/nomist} and \texttt{/parsec} flags in the \texttt{.pro} file. We opt for PARSEC over the MESA Isochrones and Stellar Tracks \citep[MIST;][]{Dotter2016} because the PARSEC models have been empirically corrected to match the observed radii of low-mass stars like \K\ \citet{Chen2014}.
    \item In a recent paper, \citet{Tayar2022} point out a persistent underestimation of uncertainties in reported stellar parameters, particularly in the wake of the high-precision parallaxes and photometric measurements from Gaia, which propagate through to computations of the stellar radius, SED, and effective temperature, and eventually trickle down to the planetary parameters we are trying to constrain. In an upcoming paper, we demonstrate how \E\ can indeed reach smaller uncertainties when performing a global fit that allows for a quasi-independent constraint on the stellar density from transit data (J.\ Eastman, \textit{in prep}). We place reasonable error floors on parameters set by the SED fit: 2.4\% on $T_\mathrm{eff}$, 2\% on $F_\mathrm{Bol}$, and 8\% on Fe/H by setting \texttt{teffsed=0.024d0}, \texttt{fehsed=0.080d0}, and \texttt{fbolsed=0.020d0} in the \texttt{.pro} file. We also manually set a conservative limit on the stellar evolution model error of 5\% in the \texttt{massradius\_parsec.pro} file (\texttt{percenterror=0.050d0} in line 260 of commit 78cc86e), which results in an acceptable uncertainty on the stellar mass of 5\%. This inflation of the various error floors results in uncertainties in the final stellar parameters that agree with systematic error floors recommended by \citet{Tayar2022}, while still allowing \E\ to use available information to best constrain the free parameters.
    \item \E\ returns a negative value for the median mass of \K d, which is not physical. The \citet{Chen2017} mass prediction for \K d's radius is $3.00^{+2.15}_{-1.02}$ M$_\oplus$, which would produce a signal of $K\sim80$ cm s\tsup{-1}, within the known systematics of the precision radial velocity instruments used in this study. We note however that the \citet{Chen2017} radius-mass relation has a number of detection biases. In cases where we wish to use an estimated value for \K d's mass we follow non-parametric approach to compute masses from radii \citep{Ning2018}, applied to an M-dwarf sample \citet{Kanodia2019}, which returns an estimate of $1.89^{+3.36}_{-0.91} M_\oplus$. This more conservative mass estimate for \K d lies even farther below the systematic noise floor of the radial velocity data.
\end{itemize}

As a check for the results of the \E\ fit we calculate the mass of the star \K\ using the mass-luminosity-metallicity relation for low-mass stars derived by \citet{Mann2019} using nearby binaries. By providing a star's K-band magnitude, distance, and metallicity, the associated code\footnote{\href{https://github.com/awmann/M\_-M\_K-}{github.com/awmann/M\_-M\_K-}} will provide a stellar mass to 2-3\% uncertainty. Providing these values for \K\ we get a result of $0.551\pm0.014$ R$_{\odot}$ from the \citet{Mann2019} relation, which is in excellent agreement with the stellar mass we find from the \E\ fit (Table~\ref{tab:exofastvalues}).

\begin{figure}
\includegraphics[width=0.48\textwidth]{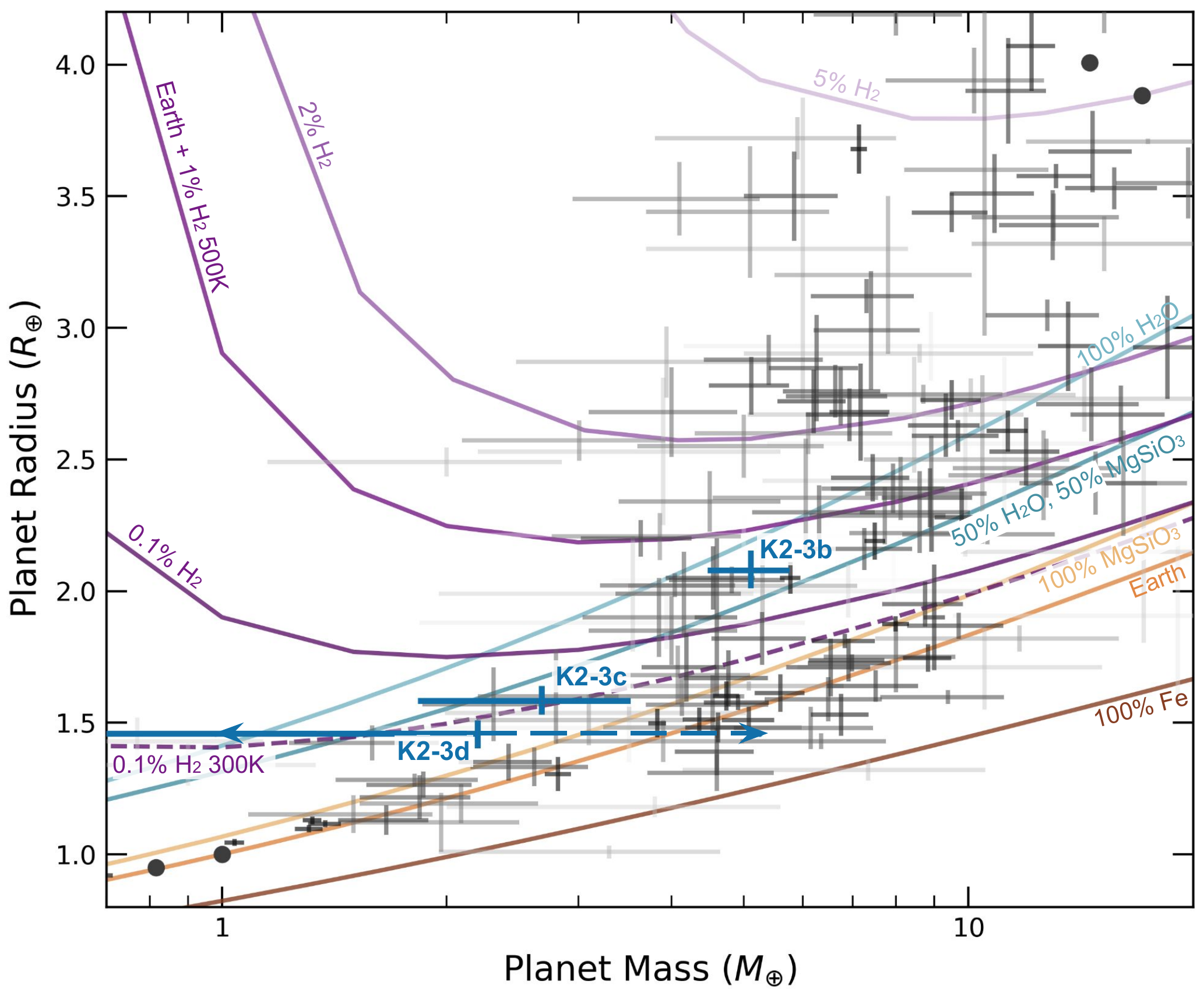}
\caption{Mass-radius diagram of small planets with measured radii and masses. Black circles are Solar System planets. Radii and masses of the \K\ planets are from Table~\ref{tab:exofastvalues}. \K d does not have a measured mass; we plot \K d's measured radius at the position of the 3$\sigma$ upper mass limit ($2.2$ M$_\oplus$). With a dashed line and arrows we also mark the estimate of \K d's mass from a non-parametric fit to a sample of M dwarfs \citep{Kanodia2019}. Composition curves from \citet{Zeng2019}. The solid purple composition curves are for an Earth-like core surrounded by an H\tsub{2} envelope of various fractional masses at 500 K H\tsub{2}, which is approximately the equilibrium temperature of \K b. We provide one example of a similar composition curve for a 0.1\% H\tsub{2} atmosphere at 300 K, the approximate equilibrium temperature of \K d.} 
\label{fig:massradius}
\end{figure}

Our final results are consistent with those of \citet{Kosiarek2019}, with some improvement on precision (Table~\ref{tab:exofastvalues}). The increased precision on the stellar radius decreased the fractional uncertainty in the derived planetary radii from 10\% to 4\%, and increased precision on the stellar mass decreased the fractional uncertainty in the masses of \K b and c from 15\% to 13\% and from 50\% to 30\%, respectively. This increased precision on the \K\ planetary radii and masses is important to narrow the range of atmospheric models we test for each planet. The derived planetary parameters in \citet{Kosiarek2019} use stellar parameters from K2-3 that were constrained using medium-resolution ($R\sim2000$) spectra \citep{Crossfield2015}. We use the \citet{Crossfield2015} values of $T_\mathrm{eff}$ and [Fe/H] as priors in our global \E\ fit. Despite these priors, we find that \K\ has a lower effective temperature and a higher metallicity than was found by \citet{Crossfield2015}. We use the values we derive for the planetary masses and radii to place the \K\ planets in the context of other small planets with measured radii and masses (Figure~\ref{fig:massradius}).

\begin{deluxetable}{lccc}
\centering
\caption{Input data to \texttt{EXOFASTv2} for the \K\ system\label{tab:exofastdata}}
\tablewidth{0pt}
\tablehead{
\colhead{Observatory/Instrument} & \multicolumn{3}{c}{Number of Observations}
}
\startdata
\multicolumn{4}{c}{Transits}\\
\hline
                                     & $\hskip 1.5em$b$\hskip 1.5em$ & $\hskip 1.5em$c$\hskip 1.5em$ & $\hskip 1.5em$d$\hskip 1.5em$\\
K2\tsup{1}                           & $\hskip 1.5em$8$\hskip 1.5em$ & $\hskip 1.5em$4$\hskip 1.5em$ & $\hskip 1.5em$2$\hskip 1.5em$ \\
Spitzer/IRAC\tsup{2,3,4,5} 4.5$\mu$m & $\hskip 1.5em$6$\hskip 1.5em$ & $\hskip 1.5em$5$\hskip 1.5em$ & $\hskip 1.5em$4$\hskip 1.5em$\\
\hline
\multicolumn{4}{c}{Radial velocities}\\
\hline
HARPS\tsup{6}                    & \multicolumn{3}{c}{132}   \\
Magellan II/PFS\tsup{7}          & \multicolumn{3}{c}{31}     \\
HARPS-N\tsup{6,8}                & \multicolumn{3}{c}{197}   \\
Keck I/HIRES\tsup{5}             & \multicolumn{3}{c}{74}    
\enddata
\tablecomments{(1) \citet{Crossfield2015}; (2) \citet{Beichman2016}; (3)~Spitzer GO~Program~11026 (PI Werner); (4)~Spitzer GO~Program~12081 (PI Benneke); (5) \citet{Kosiarek2019}; (6) \citet{Almenara2015}; (7) \citet{Dai2016}; (8) \citet{Damasso2018}
}
\end{deluxetable}

\begin{deluxetable*}{llccc}
\centering
\caption{\K\ system parameters\label{tab:exofastvalues}}
\tablewidth{0pt}
\tablehead{
\colhead{Parameter}    & \colhead{Units}  & \multicolumn{3}{c}{Values}
}
\startdata
\multicolumn{5}{c}{Priors}\\
\hline
$T_{\mathrm{eff}}$               & Effective Temperature (K)      & \multicolumn{3}{c}{$\cal{N}$(3896, 189)} \\
$[$Fe/H$]$                       & Metallicity (dex)              & \multicolumn{3}{c}{$\cal{N}$(-0.32, 0.13)} \\
$\varpi$                         & Parallax (mas)                 & \multicolumn{3}{c}{$\cal{N}$(22.691, 0.063)} \\
$A_\mathrm{V}$                   & V-band extinction (mag)        & \multicolumn{3}{c}{$\cal{U}$(0.0, 0.09672 )}\\
\hline
\multicolumn{5}{c}{Stellar parameters}\\
\hline
RA\tsup{a}   & HH:MM:SS & \multicolumn{3}{c}{11:29:20.49} \\
Dec.\tsup{a} & DD:MM:SS & \multicolumn{3}{c}{$-$01:27:18.53} \\
$M_\mathrm{s}$                   & Mass (M$_\odot$)               & \multicolumn{3}{c}{$0.549^{+0.029}_{-0.027}$} \\
$R_\mathrm{s}$                   & Radius (R$_\odot$)             & \multicolumn{3}{c}{$0.546^{+0.018}_{-0.016}$} \\
$L_\mathrm{s}$                   & Luminosity (L$_\odot$)         & \multicolumn{3}{c}{$0.0587^{+0.0018}_{-0.0019}$} \\
$F_\mathrm{Bol}$                 & Bolometric Flux ($\times 10^{-10}$ erg cm$^{-2}$ s$^{-1}$)  & \multicolumn{3}{c}{$9.68^{+0.29}_{-0.31}$} \\
$\rho_\mathrm{s}$                & Density (g cm\tsup{-3})        & \multicolumn{3}{c}{$4.78^{+0.37}_{-0.40}$} \\
log$_{10}$($g_\mathrm{s}$)                    & Surface gravity (cm s\tsup{2}) & \multicolumn{3}{c}{$4.704^{+0.023}_{-0.026}$} \\
$T_{\mathrm{eff}}$               & Effective Temperature (K)      & \multicolumn{3}{c}{$3844^{+61}_{-63}$} \\
$[$Fe/H$]$                       & Metallicity (dex)              & \multicolumn{3}{c}{$-0.157^{+0.075}_{-0.080}$} \\
Age                              & Age (Gyr)                      & \multicolumn{3}{c}{$6.9\pm4.7$} \\
$A_\mathrm{V}$                   & V-band extinction (mag)        & \multicolumn{3}{c}{$0.042^{+0.036}_{-0.030}$} \\
$\varpi$                         & Parallax (mas)                 & \multicolumn{3}{c}{$22.689\pm0.063$} \\
$d$                              & Distance (pc)                  & \multicolumn{3}{c}{$44.07\pm0.12$}\\
\hline
\multicolumn{5}{c}{Planetary parameters}\\
\hline
 & & b & c & d\\
$P$ & Period (days) & $10.0546535^{+0.0000088}_{-0.0000091}$ & $24.646729^{+0.000044}_{-0.000042}$ & $44.55603^{0.00013}_{-0.00012}$ \\
$R_\mathrm{p}$ & Radius (R$_\oplus$) & $2.078^{+0.076}_{-0.067}$ & $1.582^{+0.057}_{-0.051}$ & $1.458^{+0.056}_{-0.051}$ \\
$M_\mathrm{p}$ & Mass (M$_\oplus$)\tsup{b} & $5.11^{+0.65}_{-0.64}$ & $2.68\pm0.85$ & $-1.1\pm1.1$ \\
$T_0$ & Time of conjunction (BJD\tsub{TDB}) & $2457165.32947\pm0.00025$ & $2457329.85688^{+0.00055}_{-0.00056}$ & $2457271.78796\pm0.00076$ \\
$a$ & Semi-major axis (AU) & $0.0747^{+0.0013}_{-0.0012}$ & $0.1357^{+0.0023}_{-0.0022}$ & $0.2014^{+0.0034}_{-0.0033}$ \\
$i$ & Inclination (deg) & $89.40^{+0.34}_{-0.22}$ & $89.84^{+0.11}_{-0.14}$ & $89.82^{+0.12}_{-0.11}$ \\
$e$ & Eccentricity & $0.107^{+0.057}_{-0.059}$ & $0.048^{+0.073}_{-0.035}$ & $0.091^{+0.120}_{-0.064}$ \\ 
$\omega$ & Argument of periastron (deg) & $188^{+32}_{-34}$ & $-180^{+100}_{-110}$ & $0^{+140}_{-150}$\\
$K$ & RV semi-amplitude (m s\tsup{-1}) & $2.27\pm0.28$ & $0.88\pm0.28$ & $-0.30^{+0.29}_{-0.30}$\\
$T_\mathrm{eq}$ & Equilibrium temperature (K)\tsup{c} & $501.3^{+5.1}_{-5.2}$ & $371.8^{+3.8}_{-3.9}$ & $305.2^{+3.1}_{-3.2}$ \\
$S$ & Instellation (S$_{\oplus}$)\tsup{c} & $10.5\pm1.0$ & $3.17\pm0.30$ & $1.44\pm0.14$ \\
$R_\mathrm{p}/R_\mathrm{s}$ & Planet-to-star radius ratio & $0.03490^{+0.00041}_{-0.00036}$ & $0.02658\pm0.00032$ & $0.02449^{+0.00041}_{-0.00040}$ \\
$a/R_\mathrm{s}$ & Scaled semi-major axis & $29.44^{+0.74}_{-0.84}$ & $53.5^{+1.3}_{-1.5}$ & $79.4^{+2.0}_{-2.3}$ \\
$b$ & Transit impact parameter & $0.31^{+0.13}_{-0.18}$ & $0.14^{+0.13}_{-0.10}$ & $0.24\pm0.16$ \\
$\rho_\mathrm{p}$ & Density (g cm\tsup{-3}) & $3.11^{+0.49}_{-0.46}$ & $3.7\pm1.2$ & $-1.9^{+1.9}_{-2.0}$ \\
log\tsub{10}($g_\mathrm{p}$) & Surface gravity (cm s\tsup{-2}) & $3.062^{+0.057}_{-0.063}$ & $3.02^{+0.12}_{-0.17}$ & $2.31^{+0.37}_{-0.57}$ \\
$e\mathrm{cos}\omega$ & & $-0.094^{+0.061}_{-0.053}$ & $-0.010^{+0.041}_{-0.085}$ & $0.00^{+0.13}_{-0.12}$ \\
$e\mathrm{sin}\omega$ & & $-0.012^{+0.041}_{-0.061}$ & $0.001\pm0.029$ & $0.003^{+0.039}_{-0.051}$ \\
\enddata
\tablecomments{Median values and 68\% confidence intervals from a global fit using \E, GitHub commit 78cc86e \citep{Eastman2019}; see Section~\ref{sec:exofast} for details. \\
\tsup{a}From Gaia DR2 \citep{GaiaMission2016,GaiaDR22018}; J2000, epoch=2015.5.\\
\tsup{b}In the case of \K d, \E\ returns a mass value of $-1.1\pm1.1$ $\mathrm{M_\oplus}$, since \K d is not detected in the RV data. This gives a 3$\sigma$ upper limit on the mass of \K d of 2.2 M$_\oplus$. Based on the measured radius for \K d, \citet{Kanodia2019} predict a mass of $1.89^{+3.36}_{-0.91}$ M$_\oplus$ when comparing \K d to other planets in their M dwarf sample.\\
\tsup{c}The equilibrium temperature of the planet is calculated according to Equation 1 of \citet{Hansen2007}: $T_{\rm eq} = T_{\rm eff}\sqrt{\frac{R_\mathrm{s}}{2a}}$, which assumes no albedo and perfect energy redistribution. The quoted statistical error is likely severely underestimated relative to the systematic error inherent in this assumption. 
}
\end{deluxetable*}

\section{Stellar Influence on Planetary atmospheres} \label{sec:models}

\subsection{Atmospheric test cases} \label{subsec:atmoscases}

We combine the high energy information from \K\ with the revised planetary parameters of \K b, c, and d in order to test a range of atmospheric cases for these worlds. We determine the possible H/He envelope mass fraction and maximum water mass fraction for \K b and c, using planet interior structure models adapted from \citet{Rogers&Seager2010a,Rogers&Seager2010b, Rogers2011}. Because there is no measured mass for \K d, we assume a terrestrial composition based on its measured radius. Assuming an Earth-composition core surrounded by a solar composition envelope, the H/He-dominated envelope around \K b would comprise  $0.36^{+0.14}_{-0.11}\%$ of the planetary mass. For \K c this value is $0.07^{+0.09}_{-0.05}\%$. Based on these mass fractions and the energy-limited mass loss rate calculated in Table~\ref{tab:highenergyvalues}, \K b would lose its primordial envelope on a timescale of $2.0^{+0.8}_{-0.6}$ Gyr, while \K c would lose its envelope on a timescale of $0.6^{+0.5}_{-0.4}$ Gyr  with a 95\% upper limit of 1.5 Gyr. Both of these timescales are shorter than the estimated age of the system (Table~\ref{tab:exofastvalues}), however see Section~\ref{subsec:systemarchitecture} for more details about potential explanations for the observed \K\ system architecture. 

Both \K b and c are also consistent with the end-member case of a 100\% H\tsub{2}O envelope. For the case of a pure water envelope surrounding an Earth-composition core, an H\tsub{2}O envelope would comprise $55^{+14}_{-12}\%$ of \K b's mass and $16^{+17}_{-10}\%$ of \K c's mass. These water mass fractions could be even higher if mixing between water and an early magma ocean is considered \citep{Dorn2021}. Since we do not have a measured mass for \K d therefore do not have any real constraints on its atmosphere, we take this opportunity to showcase the impact of \K's irradiation of \K d's atmospheric composition for several scenarios motivated by Solar System planets: we assume a terrestrial composition for \K d and focus on how the high-energy flux from \K\ would impact Earth- and Venus-like atmospheres around this world. 

Based on the compositional constraints of the \K\ planets, we investigate the following atmospheric scenarios:
\begin{itemize}
    \itemsep0em
    \item \K b: Solar-like envelope; H\tsub{2}O envelope
    \item \K c: Solar-like envelope; H\tsub{2}O envelope
    \item \K d: early Venus; modern Venus; modern Earth
\end{itemize}

\subsection{Photochemical modeling of the \K\ planets}\label{subsec:atmomodels}

To investigate the impact of the observed stellar spectrum on the atmospheres of each of the planets in the K2-3 system, we use a 1-D photochemical model, \texttt{Atmos} \citep[see][for a full description of the model]{Arney2017}, in combination with pressure-temperature profiles simulated by \texttt{HELIOS} \citep{Malik2017}. Table \ref{tab:composition} lists the major atmospheric constituents for each scenario. \texttt{Atmos} has a long heritage in simulating the atmospheres of rocky planets \citep[e.g.,][]{Kasting1983,Zahnle2008} which we use in studying the atmosphere of planet d as both a Venus- and Earth-like world. Both modern Venus and modern Earth compositions are taken from observations, while the early Venus scenario is based on simulations by \citet{Way&DelGenio2020} that specify 400 ppm of CO\tsub{2} in an N\tsub{2}-dominated atmosphere. For both early Venus and modern Earth, the tropospheric water abundance is determined from the temperature profile and an assumed relative humidity of $\sim$80\%. This underestimates the water vapor feedback but is still broadly consistent with atmospheres that have not yet entered a runaway greenhouse \citep[e.g.,][]{Kasting1988}. Additionally, the higher temperatures reduce ozone abundances in the modern Earth configuration, in line with prior work \citep{Pidhorodetska2021}.
\begin{deluxetable}{l|p{5cm}}
\centering
\caption{Compositions by Atmospheric Type \label{tab:composition}}
\tablewidth{0pt}
\tablehead{
\colhead{Type} &  \colhead{Composition}
}
\startdata
Type          & Composition \\\hline
Solar         & 85.3\% H\tsub{2}, 14.6\% He, 810 ppm H\tsub{2}O, 400 ppm CH\tsub{4}, 16 ppm CO\\\hline
Water         & 99.8\% H\tsub{2}O, 0.1\% H\tsub{2}, 996 ppm He \\\hline
Early Venus & 98.9\% N\tsub{2}, 17.7\% H\tsub{2}O, 400 ppm CO\tsub{2} \\\hline
Modern Venus  & 96.5\% CO\tsub{2}, 2.5\% N\tsub{2}, 20 ppm H\tsub{2}O \\\hline
Modern Earth         & 21\% O\tsub{2}, 15.6\% H\tsub{2}O, 360 ppm CO\tsub{2}, 1.8 ppm CH\tsub{4}, 63.4\% N\tsub{2}
\enddata
%\tablecomments{}
\end{deluxetable}

\texttt{Atmos} has also been modified to handle more diverse planetary scenarios, including warm super-Earths and sub-Neptunes \citep{Harman2022}, which covers the scenarios for planets b and c. These modifications include using higher pressure and temperature thermochemical reactions appropriate for a hydrogen-dominated atmosphere at modest instellation, as well as photolysis of key species, based on the C-H-O chemical scheme from \citet{Tsai2017,Tsai2021}. The ``surface" is set to the 10-bar pressure level. We assume a vertical eddy diffusion parameter $K_{zz} = 10^{9}$ cm$^{2}$ s$^{-1}$, noting that different values of $K_{zz}$ (within a few orders of magnitude) have small effects on the atmospheric composition \citep{Harman2022}.

\begin{figure}
\includegraphics[width=0.48\textwidth]{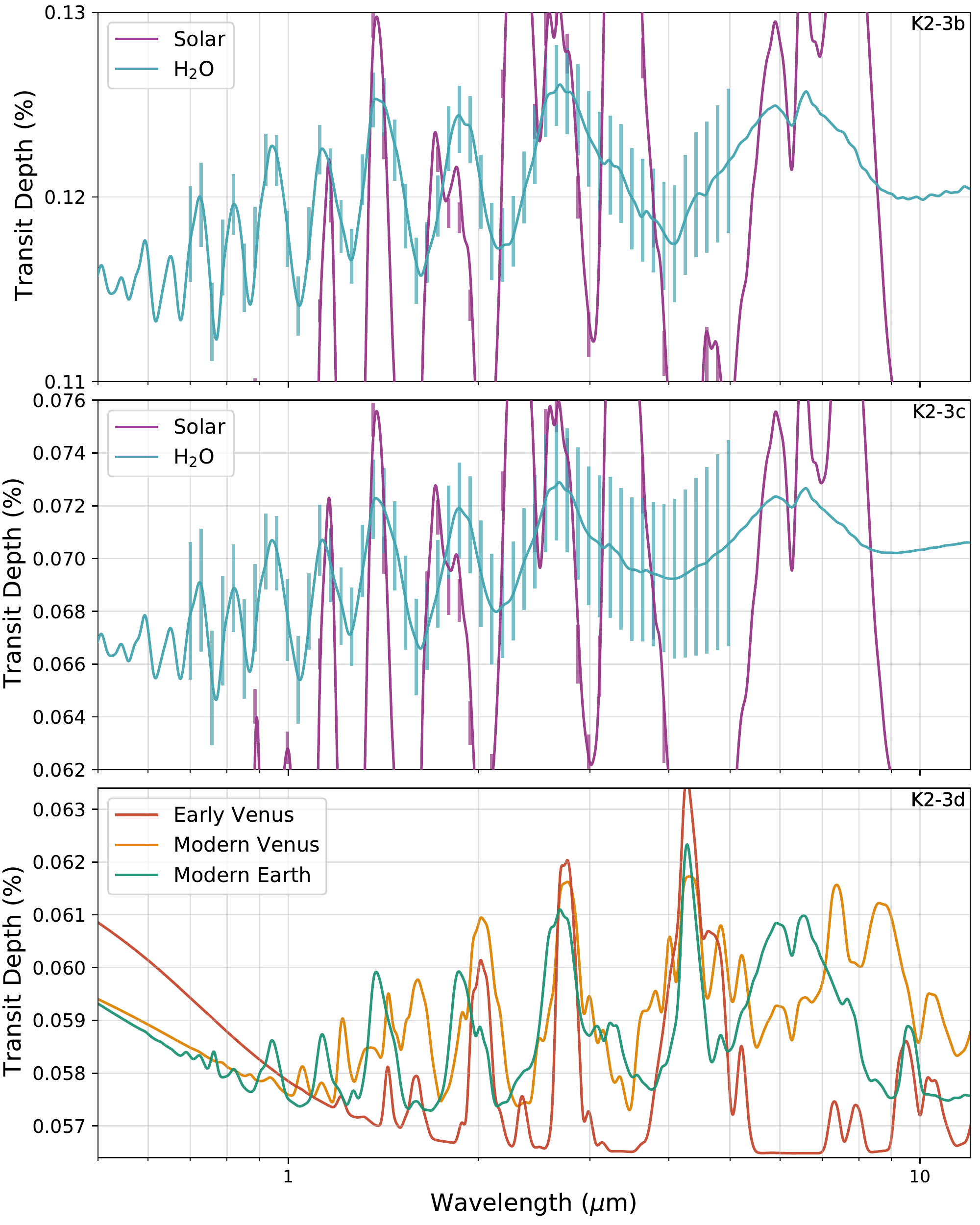}
\caption{Model transmission spectra for K2-3b, c, and d (top, middle, and bottom panels, respectively) from 0.5 to 12 $\mu$m. Error bars illustrate the typical uncertainty in measuring such atmospheres for K2-3b and c with 50 hr of in-transit observations at a resolution of 25 using JWST/NIRSpec-PRISM. Error
bars are not shown for the model atmosphere cases on K2-3d since the uncertainties of 2.7 ppm are much larger than the model atmosphere features. Note the different y-axis scales across the three panels.} 
\label{fig:spectra}
\end{figure}

The results of the photochemical simulations are passed to the Planetary Spectrum Generator \citep[PSG;][]{Villanueva2018} to produce model transmission spectra. With 50 hours of in-transit observing time with JWST/NIRSpec we would achieve a typical uncertainty of 2.7 ppm at a resolution of 25 (Figure \ref{fig:spectra}. The solar composition and H\tsub{2}O-dominated cases for \K b and c are readily differentiated from one another with these simulated observations. For the \K d atmospheric cases the 2.7 ppm uncertainty is larger than the total spectral variation (Figure~\ref{fig:spectra}; features on the order of 1 ppm). In other words, 50 hours of in-transit observations of \K d with JWST/NIRSpec could not distinguish between the three atmospheric cases we explore for this planet. We use the 3$\sigma$ observational upper limit for the mass of \K d (2.2 M$_\oplus$) when generating the atmospheric models. We do not include the behavior of aerosols (clouds or hazes) in the model transmission spectra.
%50 hrs in-transit time = 20, 14, 12 transits for b, c, d

To test the influence of the high-energy stellar spectrum on the planetary atmospheres, we run the models again with the input stellar spectrum increased and decreased by a factor of 10. An order of magnitude is about the spread in emission features between similar M dwarfs \citep{France2016}, so with this simple experiment we attempt to address the impact of using a measured high-energy stellar spectrum versus a proxy star. We outline the results for each of the planetary atmospheres:

\begin{itemize}
    \item For \K b there is no change in the volume mixing ratio of prominent molecules like water in either the solar composition or H\tsub{2}O-dominated cases. This is because at an equilibrium temperature of about 500 K, it is the thermochemistry that dominates the reaction rates in the upper atmosphere.
    \item For \K c, at an equilibrium temperature of about 370 K, the reactions driven by photolysis are on par with those driven by thermochemistry, allowing us to see differences in trace species volume mixing ratios in the solar composition case. In the pure H\tsub{2}O atmosphere case we do not see any change.
    \item For \K d we test a more complex set of atmospheres and indeed we see a more complicated set of results. Elemental sulfur (S\tsub{8}), for example, decreases by orders of magnitude as the UV flux increases, implying that for warm to temperate terrestrial planets, a magnitude of uncertainty in the UV will play a significant role in the chemistry of the atmosphere. Molecular oxygen (O\tsub{2}) also changes, with volume mixing ratio increasing by orders of magnitude in the 10$\times$ more UV flux case compared to the 10$\times$ less UV flux. This demonstrates that under certain UV or T/P conditions on terrestrial worlds, O\tsub{2} can arise as an abiotic false positive.
\end{itemize}

\section{Discussion}\label{sec:discussion}

\subsection{Explaining the K2-3 system architecture}\label{subsec:systemarchitecture}

The \K\ system is an early-M star orbited by, in increasing orbital distance, a sub-Neptune, a super-Earth enriched in volatiles, and a likely terrestrial super-Earth. It should be noted here that multi-planet systems consisting of small planets are less likely to have a mixture of sub-Neptunes and super-Earths than just one type or the other \citep{Millholland2021}. We place the \K\ planets in the context of the small planet radius valley, extended to a second dimension of planetary instellation (or, similarly, orbital period). When looking at the radius valley for FGK stars \citep[e.g., Figure 10 in][]{Fulton2017}, there is a trend in the location of the radius valley towards increasing planet radius with increasing instellation. When reproducing this figure for stars with $T_\mathrm{eff}<4700$ K (spectral type K3.5V and later), the trend in the location of the radius valley appears to flip, and increases in planet radius with \textit{decreasing} instellation \citep[e.g., Figure 11 of ][]{Cloutier2020a}.

\begin{figure}
\includegraphics[width=0.48\textwidth]{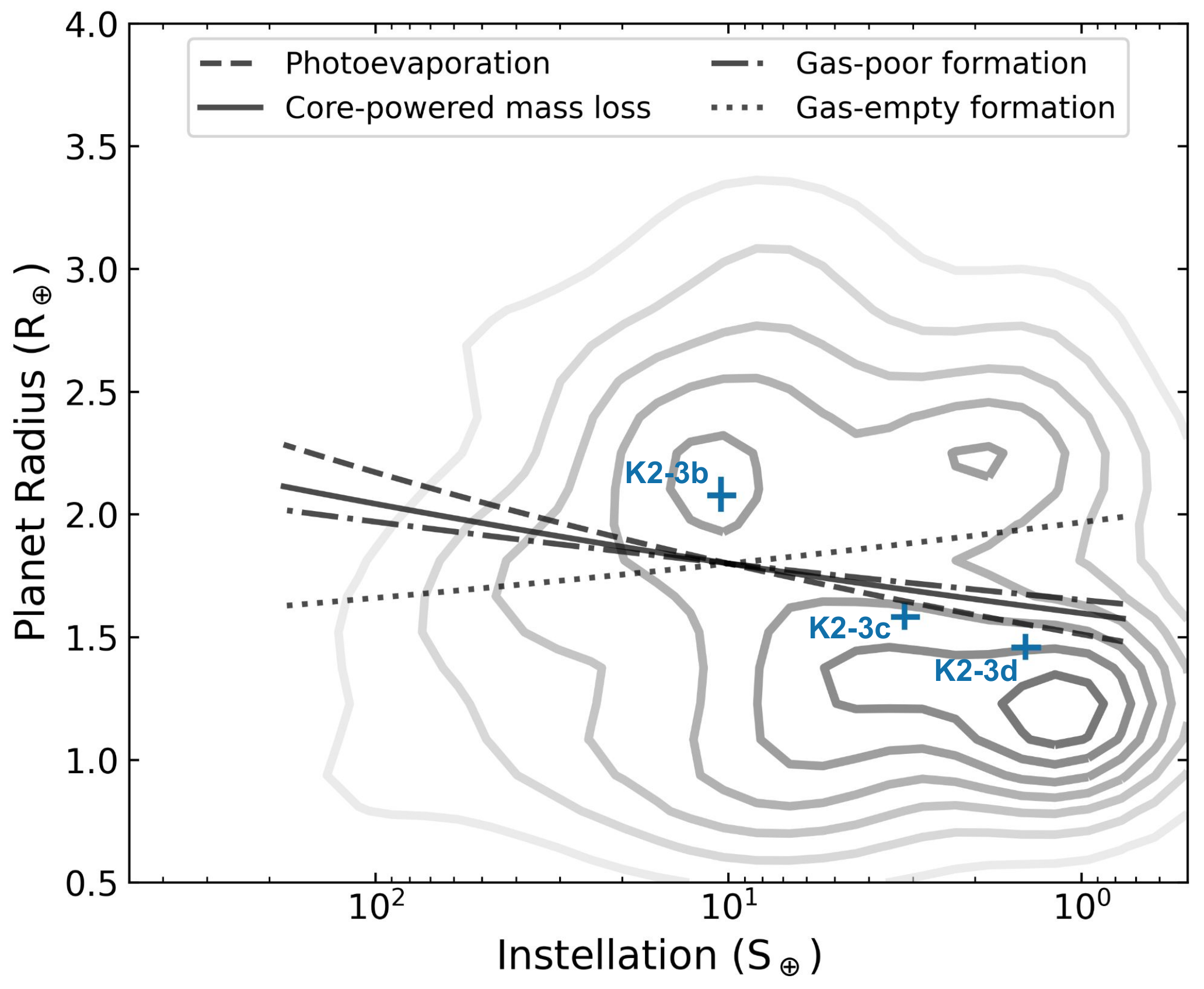}
\caption{Small planet occurrence rate for planets orbiting cool stars ($T_\mathrm{eff}<4700$ K) in radius-instellation space. Occurrence contours are from \citet{Cloutier2020a}. Curves represent trends in the radius valley gap as a function of instellation. The trends in $d\ \mathrm{log}\ R_\mathrm{p}/d\ \mathrm{log}\ S$ and their sources are as follows: Photoevaporation \citep[0.11;][]{Lopez2018}, core-powered mass loss \citep[0.08;][]{Gupta&Schlichting2019}, gas-poor formation \citep[0.06;][]{Lee2021}, gas-empty formation \citep[-0.08;][]{Lopez2018}. The \K\ planets are labeled in blue.} 
\label{fig:popcontours}
\end{figure}

As can be seen in Figure~\ref{fig:popcontours}, the \K\ planets appear to fit the bi-modal small planet distribution, with \K b falling in the sub-Neptune category and \K c and d falling into the super-Earth category, however we should look more closely at these planets together as members of the same multi-planet system to further test theories of planetary formation and evolution. The lines in Figure~\ref{fig:popcontours} demonstrate how different planetary formation and evolutionary theories produce trends in the small planet radius valley with planetary instellation that separate the populations of sub-Neptunes from the super-Earths, though the current planet population around low mass stars is not yet large enough to definitively favor one mechanism over the other \citep{Cloutier2020a}. Focusing on the \K\ planets together we ask: \textit{can planetary formation and evolution models produce the \K\ system architecture?} To answer this question we address each theory below.

\textit{Photoevaporation}: We use the publicly available photoevaporation code \texttt{EvapMass} \citep{Owen2020a} to test whether pairs of sub-Neptune and super-Earth planets in the \K\ system are consistent with photoevaporation, a process by which high-energy stellar flux drives atmospheric mass loss. This process is most effective during the first 100 Myr of a planet's lifetime when the star is most active. Photoevaporation rates depend on the amount of high-energy XUV stellar flux a planet receives. Because the \K\ planets orbit the same star, we can take this value out of the equation when comparing the \K\ planets to each other. We start with the \K b and d pair, since \K b is a sub-Neptune and \K d is potentially a terrestrial world given its radius \citep{Chen2017,Kanodia2019}. Using stellar and planetary parameters from Table~\ref{tab:exofastvalues}, \texttt{EvapMass} calculates the minimum mass of the sub-Neptune \K b that would be consistent with photoevaporation if this method were responsible for shaping both \K b and \K d.

\texttt{EvapMass} cannot find a minimum mass for the sub-Neptune \K b assuming that the super-Earth \K d has just lost a primordial atmosphere. Put simply, the \K d--\K b pair are not consistent with photoevaporation. If photoevaporation were solely responsible for the \K\ system architecture, it would require some combination of \K b having a much larger or denser core, or \K d orbiting much closer to the star, neither of which is the case. This may not be surprising since the sub-Neptune \K b currently receives 7 times more XUV flux than \K d (Table~\ref{tab:highenergyvalues}). \texttt{EvapMass} also does not return a valid result for the \K b and c pair, here assuming that \K c is the terrestrial world. (This is not a great assumption since \K c falls right in the valley between the super-Earth and sub-Neptune distributions, and it's mass and radius point to volatile enrichment above what is expected for rocky planets.) \citet{Owen2020a} explored 73 multi-planet Kepler systems with \texttt{EvapMass} and find only two to be inconsistent with photoevaporation. In both cases the system architecture has a sub-Neptune on an orbit between an inner and outer terrestrial planet. In the \K\ system architecture the sub-Neptune is the inner-most of three planets. It should be noted that there are cases in the Kepler system where a sub-Neptune orbits interior to a super-Earth and the system is still consistent with photoevaporation. We note some assumptions that go into \texttt{EvapMass} in more generally into photoevaporation theory. One is that the planetary cores are all Earth-like surrounded by a $\gtrsim$1\% H/He envelope accreted from the protoplanetary disc. As discussed in Section~\ref{subsec:atmoscases}, the solar composition envelopes for \K b and c are disfavored. If instead their atmospheres contain some amount of H\tsub{2}O they could be more resilient to photoevaporation. 

We compare the result from \texttt{EvapMass} to similar calculations aimed at reconciling multi-planet systems with photoevaporation. In Appendix A of \citet{Cloutier2020b} the authors lay out a set of inequalities to determine the minimum mass of a gaseous sub-Neptune (or maximum mass of a rocky super-Earth) to satisfy the photoevaporation model. Because inequalities depend on both masses and radii, we test a grid of models that encompasses the range of possible  masses and radii for \K d (we use the range of masses predicted by \citet{Kanodia2019}). We keep the mass and radius of \K b, along with semi-major axes of both planets, fixed to the median values from Table~\ref{tab:exofastvalues}. We find that when taking into account the uncertainty in the radius and mass of \K d, \K b--\K d pairings in which \K d has a bulk density less than about 3.5 g cm\tsup{-3} are consistent with photoevaporation. This would mean that \K d would have to have some volatile enrichment above what is expected for terrestrial worlds of with similar radii. However we note that, unlike the \texttt{EvapMass} code, the \citet{Cloutier2020b} inequalities do not take into account contraction of the primordial envelope over time as material is lost.

\textit{Core-powered mass loss}: In principle, core-powered mass loss, by which a light H/He envelope can be driven off by energy originating in the planetary core, might explain the \K\ system architecture. Since the driving energy is left over from planetary formation, a stochastic process, it is possible that \K c and d lost most or all of their H/He envelopes while \K b was able to maintain one. However, the mechanism of core-powered mass loss relies on energy from the core reaching the outer isothermal layer of a light atmosphere \citep[an atmosphere weighing less than 5\% of the planet's core mass;][]{Ginzburg2016}, but the rate at which this happens depends on the atmospheric sound speed $c_s$, which is itself a function of the planet's equilibrium temperature \citep{Gupta&Schlichting2019}. In this way, core-powered mass loss takes into account the distance of the planet from the star. Thus, if we make the assumption that the \K\ planets were born with identical H/He envelopes, the core-powered mass loss mechanism also disfavors the observed system architecture.

\citet{Cloutier2020b} also provide an inequality that must be satisfied for a multi-planet system to be a viable outcome of core-powered mass loss (appendix B of that paper). Based on equations laid out in \citet{Ginzburg2018,Gupta&Schlichting2019}, it is necessary for the cooling timescale of a sub-Neptune planet (like \K b) to be greater than the cooling timescale of a super-Earth (like \K d) in the same system. As with the photoevaporation case, we test a grid of inequalities corresponding to the range of possible masses and radii for \K d. Given the dependence of the cooling timescales on equilibrium temperature, we again find that if \K b and d both formed with H/He envelopes, their current architecture cannot be easily explained solely by core-powered mass loss. \K b and \K d pairings that satisfy the core-powered mass loss inequality would give \K d a lower bulk density than expected for a terrestrial world.

\textit{Gas-poor formation}: Another explanation of the small planet radius valley is that this distribution is baked in at formation. In this picture, the inner region of a protoplanetary disk becomes depleted in gas towards the end of the disk lifetime. At this point protoplanetary cores in the inner disk can cross orbits, collide, and grow. Small amounts of gas escape from the gas-rich outer disk to feed the gas-poor inner disk, which can be acquired by the growing cores \citep{Lee&Chiang2016}. In this way sub-Neptunes and super-Earths can form \textit{in situ}. Late-time gas accretion alone can reproduce the small planet radius valley for FGK stars, with super-Earths favored closer to their stars \citep{Lee2021}. Post-processing by photoevaporation and/or core-powered mass loss can further turn some of the sub-Neptune population into super-Earths, improving even more the fit to the small planet radius valley for FGK stars. So, while gas-poor formation can reproduce the small planet radius valley, it does not inherently allow for the picture of a sub-Neptune orbiting interior to super-Earths in the same system. It it not certain how this picture would change for planetary formation in M dwarf discs. Recent planetary formation work demonstrates that super-Earth-sized planets can indeed be formed at periods $>$30 d (\K b has a period of 44 d), where the influence of photoevaporation and core-powered mass loss is diminished \citep{Lee2022}. However this still does not easily explain the sub-Neptune \K b orbiting well interior to \K d.

A more extreme formation scenario would be gas-empty formation, in which there is truly no access of late-forming rocky cores to gas from the disc \citep{Lopez2018}. Though this can be considered more of a toy model, it does more accurately fit the tentative trend in the small planet radius valley for low-mass stars (Figure~\ref{fig:popcontours}). However, this model would still have difficulty explaining the \K\ system architecture because of the model dependency on incident stellar flux, which is a function of the planet orbital distance.

\textit{Impact erosion}: Finally, we can also consider impact erosion, whereby the collision of material (e.g., asteroids, comets, planetessimals) with a planet surface can cause its atmosphere to either grow or deplete. Whether the impacted planet's atmosphere grows or depletes is the result of a balance of factors, including the speed, size, impact angle, and volatile content of the impactor, as well as the mass of the impacted planet's atmosphere and escape velocity. However, when calculating the effects of impact erosion on terrestrial planets, \citet{Wyatt2020} include a dependence on the planet's orbital location: the erosional efficiency of an impact includes assumptions about the atmospheric scale height, which in turn depends on the planet's semi-major axis. This results in the conclusion that, all else being equal, terrestrial planets closer to their stars are more likely to lose their atmospheres after an impact than those farther away. \citet{Wyatt2020} do not consider planets with thick H/He envelopes, but the general conclusion does not line up with the observed \K\ system architecture.

An extreme variation on impact erosion is a late-stage giant impact, whereby the impact can lead to atmospheric loss not only at the site of the impact but also globally due to shocks propagating through the planet \citep{Schlichting2015}. Though close-in planets are again more susceptible to giant impacts because the extent of their isothermal outer envelopes depends on their equilibrium temperatures \citep{Biersteker&Schlichting2019}, the stochasticity of the giant impact process makes for a tempting, though somewhat fine-tuned, explanation for the \K\ system: perhaps \K c and d started out more like sub-Neptune worlds, only to suffer late-stage giant impacts in which most of their low mean molecular weight envelopes were removed. This might leave \K c and d mostly rocky but enhanced with volatile material, as observed.

\vspace{.25cm}

Finally, we address the fact that the EUV estimate of \K\ and the resulting mass loss rates depend on our assumptions about the X-ray emission from \K, for which we only have an upper limit. As a rough check, we estimate the X-ray flux from \K\ by translating its rotation period \citep[40 d;][]{Kosiarek2019} into an X-ray luminosity using the relations from \citet{Wright2018}. The result is an estimated X-ray flux of $2.35\times10^{-15}$ erg cm\tsup{-2} s\tsup{-1}, almost an entire order of magnitude less than our assumed upper limit. Propagating this value through to the estimated EUV flux of \K, and further onto the mass loss rates of the planets \K b and c, we find that these worlds would lose their H/He envelopes in approximately 5.6 and 1.8 Gyr, respectively. While this would mean that \K b may indeed be in the process of losing its primordial atmosphere and will eventually turn into a stripped core, the 1.8 Gyr timescale of hydrodynamic mass loss for \K c is at the lowest end of the estimated age of \K, making it unlikely that \K c is still in the process of losing a primordial atmosphere. We are therefore skeptical that photoevaporation or core powered mass loss can explain the overall architecture of the \K\ system.

\subsection{Prospects for habitable conditions on K2-3d}

The outer-most (known) planet of the \K\ system has a terrestrial-like radius but no measured mass, and sits at the inner edge of \K's habitable zone \citep{Kopparapu2014}. If \K d is relatively water-poor, it may have clement surface conditions and liquid water near the poles \citep[e.g.,][]{Abe2011}. Alternatively, as a synchronously rotating planet, \K d could feature a thick sub-stellar cloud deck that would extend the inner edge of the habitable zone to higher instellation \citep[e.g.,][]{Yang2013,Kopparapu2017}. However, since \K d receives roughly 1.4 times Earth's insolation, it could be in the moist greenhouse regime \citep{Kopparapu2017}, although the threshold for this regime varies based on a number of planetary parameters, including whether or not there is a surface ocean to efficiently distribute heat around the planet \citep{Yang2019}.

\K d may be further enriched in volatiles, and in particular water, than the compositions of \citet{Zeng2019} imply (Figure~\ref{fig:massradius}). For instance, if \K d did manage to accrete H/He from the protoplanetary nebula, surface pressures would be high enough to support a surface magma ocean into which water would readily mix. This would lead to an increase in the estimated water mass fraction given the observed planet radius, compared to the simpler picture of a silicate rock surrounded by a layer of volatiles \citep{Dorn2021}. This presents an intriguing picture for habitability for \K d, which could have water locked up in its mantle. Having lost its H/He envelope, or never having had one in the first place, we speculate that this could lead \K d to possess a water-rich secondary atmosphere. However, such an atmosphere would likely transition into a runaway greenhouse state due to the high instellation, increasing the apparent volatile abundance over its actual inventory \citep[e.g.,][]{Turbet2019}. If, for example, \K d has an Earth-like water inventory, the higher XUV fluxes from \K\ would drive rapid atmospheric escape and evolve the planet towards a modern Venus-like state (Figure \ref{fig:spectra}).

\section{Conclusion} \label{sec:conclusion}

In this work we investigate the \K\ system using a blend of observations and theory. Because the \K\ planets span the radius-mass space of small exoplanets we can use these worlds as testing grounds for atmospheric and planet evolution models over a range of instellations and compositional cases. 

We measure the high-energy spectrum of \K\ directly using HST/COS in the UV and XMM-Newton in the X-ray. We use this information to estimate regions of \K's high energy spectrum that we cannot measure directly, such as the \Lya\ line and the EUV, via the differential emission measure method \citep{Duvvuri2021}. We combine the high-energy measurements and estimates with a BT-Settl stellar atmosphere model \citep{Allard2014} in order to construct a panchromatic spectrum of \K\ that can be passed to atmospheric models. 

We use all available radial velocity and transit data of the \K\ planets along with precise parallaxes from Gaia in order to perform a joint fit using the \E\ package. The result yields precise radius measurements of the \K\ planets down to 4\% precision. The masses remain more uncertain. The masses of \K b and c are constrained to 13\% and 30\%, respectively, while the outermost world \K d is not detected in the radial velocity measurements so we estimate its mass using empirical radius-mass correlations from \citet{Chen2017}. These radius and mass measurements allow us to constrain the maximum and minimum water/hydrogen envelope mass fractions for \K b and c, which both show evidence of volatile-enhanced compositions relative to a terrestrial Earth-like composition (Figure~\ref{fig:massradius}). 

We model several potential atmospheric compositions for all three of the known planets in the \K\ system, focusing on solar composition and pure-water atmospheres for the two innermost planets, b and c, and exploring three rocky planet atmosphere scenarios for planet d---early Venus, modern Venus, and modern Earth. \K d is located on the cusp of the inner edge of the habitable zone, and could possess an atmosphere that is water-rich or water-poor, depending on how the planet and host star evolved. If \K d did accrete an H/He envelope from the protoplanetary nebula before losing it, it may have locked up a substantial amount of water in its mantle during a prolonged magma ocean phase, which could lead to a Venus-like atmosphere today.

Finally, while the \K\ planets as individual worlds fit nicely into the bi-modal distribution of small planet radii, when scrutinized as a whole system the \K\ planet properties present challenges to current theories of planet formation and evolution. We find that while \K b can maintain a H/He envelope in the presence of host star \K, the overall system architecture does not easily fit with the photoevaporation and core-powered mass loss mechanisms to explain the super-Earth--sub-Neptune radius valley. Given that we do not have a measured mass for \K d, and because \K c has a terrestrial-like radius but is enhanced in volatiles, it is possible that this system is exhibiting the outcomes of the more stochastic processes of planet formation and evolution. We posit that the three \K\ planets are enhanced in volatiles as an outcome of the planet formation process, subsequent planet migration, post-formation delivery of volatile material, or impact erosion.

\vspace{.5cm}
%\begin{acknowledgments}
We thank Ryan Cloutier, Beatriz Campos Estrada, Jo\~{a}o Mendon\c{c}a, Victoria Antoci, Ren\'e Tronsgaard Rasmussen, Evgenya Shkolnik, Amber Medina, and Zoe Oliver-Grey for helpful discussions during the preparation of this manuscript. This work is based on observations made with the NASA/ESA Hubble Space Telescope, obtained at the Space Telescope Science Institute. Support for program \#15110 was provided by NASA through a grant from the Space Telescope Science Institute, which is operated by the Association of Universities for Research in Astronomy, Inc., under NASA contract NAS5-26555. This work has made use of data from the European Space Agency (ESA) mission {\it Gaia} (\url{https://www.cosmos.esa.int/gaia}), processed by the {\it Gaia} Data Processing and Analysis Consortium (DPAC, \url{https://www.cosmos.esa.int/web/gaia/dpac/consortium}). Funding for the DPAC has been provided by national institutions, in particular the institutions participating in the {\it Gaia} Multilateral Agreement. This work makes use of the CHIANTI atomic database. CHIANTI is a collaborative project involving George Mason University, the University of Michigan (USA), University of Cambridge (UK) and NASA Goddard Space Flight Center (USA). This research has made use of the NASA Exoplanet Archive, which is operated by the California Institute of Technology, under contract with the National Aeronautics and Space Administration under the Exoplanet Exploration Program.
%\end{acknowledgments}

\facilities{HST (COS), 
            XMM (EPIC pn), 
            Spitzer (IRAC), 
            Kepler (K2),
            Magellan:Clay (Planet Finder Spectrograph),
            ESO:3.6m (HARPS),
            TNG (HARPS-N),
            Keck:I (HIRES),
            Exoplanet Archive
            }

\software{\texttt{EvapMass} \citep{Owen2020b}
          \texttt{exoatlas} (\href{https://github.com/zkbt/exoplanet-atlas}{github.com/zkbt/exoplanet-atlas}),
          \texttt{EXOFASTv2} \citep{Eastman2017}, 
          \texttt{keplersplinev2} (\href{https://github.com/avanderburg/keplersplinev2}{github.com/avanderburg/keplersplinev2}),
          \texttt{mk-mass} \citep{Mann2019}
         }

Data products are available as HLSPs at MAST: \dataset[10.17909/t9-fqky-7k61]{\doi{10.17909/t9-fqky-7k61}}.

\bibliography{MasterBibliography}

\begin{thebibliography}{}
\expandafter\ifx\csname natexlab\endcsname\relax\def\natexlab#1{#1}\fi
\providecommand{\url}[1]{\href{#1}{#1}}
\providecommand{\dodoi}[1]{doi:~\href{http://doi.org/#1}{\nolinkurl{#1}}}
\providecommand{\doeprint}[1]{\href{http://ascl.net/#1}{\nolinkurl{http://ascl.net/#1}}}
\providecommand{\doarXiv}[1]{\href{https://arxiv.org/abs/#1}{\nolinkurl{https://arxiv.org/abs/#1}}}

\bibitem[{Abe {et~al.}(2011)Abe, Abe-Ouchi, Sleep, \& Zahnle}]{Abe2011}
Abe, Y., Abe-Ouchi, A., Sleep, N.~H., \& Zahnle, K.~J. 2011, Astrobio., 11,
  443, \dodoi{10.1089/ast.2010.0545}

\bibitem[{{Allard}(2014)}]{Allard2014}
{Allard}, F. 2014, in Exploring the Formation and Evolution of Planetary
  Systems, ed. M.~{Booth}, B.~C. {Matthews}, \& J.~R. {Graham}, Vol. 299,
  271--272, \dodoi{10.1017/S1743921313008545}

\bibitem[{{Almenara} {et~al.}(2015){Almenara}, {Astudillo-Defru}, {Bonfils},
  {Forveille}, {Santerne}, {Albrecht}, {Barros}, {Bouchy}, {Delfosse},
  {Demangeon}, {Diaz}, {H{\'e}brard}, {Mayor}, {Neves}, {Rojo}, {Santos}, \&
  {W{\"u}nsche}}]{Almenara2015}
{Almenara}, J.~M., {Astudillo-Defru}, N., {Bonfils}, X., {et~al.} 2015, \aap,
  581, L7, \dodoi{10.1051/0004-6361/201525918}

\bibitem[{{Arney} {et~al.}(2017){Arney}, {Meadows}, {Domagal-Goldman},
  {Deming}, {Robinson}, {Tovar}, {Wolf}, \& {Schwieterman}}]{Arney2017}
{Arney}, G.~N., {Meadows}, V.~S., {Domagal-Goldman}, S.~D., {et~al.} 2017,
  \apj, 836, 49, \dodoi{10.3847/1538-4357/836/1/49}

\bibitem[{{Beichman} {et~al.}(2016){Beichman}, {Livingston}, {Werner},
  {Gorjian}, {Krick}, {Deck}, {Knutson}, {Wong}, {Petigura}, {Christiansen},
  {Ciardi}, {Greene}, {Schlieder}, {Line}, {Crossfield}, {Howard}, \&
  {Sinukoff}}]{Beichman2016}
{Beichman}, C., {Livingston}, J., {Werner}, M., {et~al.} 2016, \apj, 822, 39,
  \dodoi{10.3847/0004-637X/822/1/39}

\bibitem[{{Biersteker} \& {Schlichting}(2019)}]{Biersteker&Schlichting2019}
{Biersteker}, J.~B., \& {Schlichting}, H.~E. 2019, \mnras, 485, 4454,
  \dodoi{10.1093/mnras/stz738}

\bibitem[{{Bressan} {et~al.}(2012){Bressan}, {Marigo}, {Girardi}, {Salasnich},
  {Dal Cero}, {Rubele}, \& {Nanni}}]{Bressan2012}
{Bressan}, A., {Marigo}, P., {Girardi}, L., {et~al.} 2012, \mnras, 427, 127,
  \dodoi{10.1111/j.1365-2966.2012.21948.x}

\bibitem[{{Chen} \& {Kipping}(2017)}]{Chen2017}
{Chen}, J., \& {Kipping}, D. 2017, \apj, 834, 17,
  \dodoi{10.3847/1538-4357/834/1/17}

\bibitem[{{Chen} {et~al.}(2014){Chen}, {Girardi}, {Bressan}, {Marigo},
  {Barbieri}, \& {Kong}}]{Chen2014}
{Chen}, Y., {Girardi}, L., {Bressan}, A., {et~al.} 2014, \mnras, 444, 2525,
  \dodoi{10.1093/mnras/stu1605}

\bibitem[{{Claret} \& {Bloemen}(2011)}]{Claret2011}
{Claret}, A., \& {Bloemen}, S. 2011, \aap, 529, A75,
  \dodoi{10.1051/0004-6361/201116451}

\bibitem[{{Cloutier} \& {Menou}(2020)}]{Cloutier2020a}
{Cloutier}, R., \& {Menou}, K. 2020, \aj, 159, 211,
  \dodoi{10.3847/1538-3881/ab8237}

\bibitem[{{Cloutier} {et~al.}(2020){Cloutier}, {Eastman}, {Rodriguez},
  {Astudillo-Defru}, {Bonfils}, {Mortier}, {Watson}, {Stalport}, {Pinamonti},
  {Lienhard}, {Harutyunyan}, {Damasso}, {Latham}, {Collins}, {Massey}, {Irwin},
  {Winters}, {Charbonneau}, {Ziegler}, {Matthews}, {Crossfield}, {Kreidberg},
  {Quinn}, {Ricker}, {Vanderspek}, {Seager}, {Winn}, {Jenkins}, {Vezie},
  {Udry}, {Twicken}, {Tenenbaum}, {Sozzetti}, {S{\'e}gransan}, {Schlieder},
  {Sasselov}, {Santos}, {Rice}, {Rackham}, {Poretti}, {Piotto}, {Phillips},
  {Pepe}, {Molinari}, {Mignon}, {Micela}, {Melo}, {de Medeiros}, {Mayor},
  {Matson}, {Martinez Fiorenzano}, {Mann}, {Magazz{\'u}}, {Lovis},
  {L{\'o}pez-Morales}, {Lopez}, {Lissauer}, {L{\'e}pine}, {Law}, {Kielkopf},
  {Johnson}, {Jensen}, {Howell}, {Gonzales}, {Ghedina}, {Forveille},
  {Figueira}, {Dumusque}, {Dressing}, {Doyon}, {D{\'\i}az}, {Fabrizio},
  {Delfosse}, {Cosentino}, {Conti}, {Collins}, {Cameron}, {Ciardi}, {Caldwell},
  {Burke}, {Buchhave}, {Brice{\~n}o}, {Boyd}, {Bouchy}, {Beichman}, {Artigau},
  \& {Almenara}}]{Cloutier2020b}
{Cloutier}, R., {Eastman}, J.~D., {Rodriguez}, J.~E., {et~al.} 2020, \aj, 160,
  3, \dodoi{10.3847/1538-3881/ab91c2}

\bibitem[{{Crossfield} {et~al.}(2015){Crossfield}, {Petigura}, {Schlieder},
  {Howard}, {Fulton}, {Aller}, {Ciardi}, {L{\'e}pine}, {Barclay}, {de Pater},
  {de Kleer}, {Quintana}, {Christiansen}, {Schlafly}, {Kaltenegger}, {Crepp},
  {Henning}, {Obermeier}, {Deacon}, {Weiss}, {Isaacson}, {Hansen}, {Liu},
  {Greene}, {Howell}, {Barman}, \& {Mordasini}}]{Crossfield2015}
{Crossfield}, I.~J.~M., {Petigura}, E., {Schlieder}, J.~E., {et~al.} 2015,
  \apj, 804, 10, \dodoi{10.1088/0004-637X/804/1/10}

\bibitem[{{Dai} {et~al.}(2016){Dai}, {Winn}, {Albrecht}, {Arriagada},
  {Bieryla}, {Butler}, {Crane}, {Hirano}, {Johnson}, {Kiilerich}, {Latham},
  {Narita}, {Nowak}, {Palle}, {Ribas}, {Rogers}, {Sanchis-Ojeda}, {Shectman},
  {Teske}, {Thompson}, {Van Eylen}, {Vanderburg}, {Wittenmyer}, \&
  {Yu}}]{Dai2016}
{Dai}, F., {Winn}, J.~N., {Albrecht}, S., {et~al.} 2016, \apj, 823, 115,
  \dodoi{10.3847/0004-637X/823/2/115}

\bibitem[{{Damasso} {et~al.}(2018){Damasso}, {Bonomo}, {Astudillo-Defru},
  {Bonfils}, {Malavolta}, {Sozzetti}, {Lopez}, {Zeng}, {Haywood}, {Irwin},
  {Mortier}, {Vanderburg}, {Maldonado}, {Lanza}, {Affer}, {Almenara},
  {Benatti}, {Biazzo}, {Bignamini}, {Borsa}, {Bouchy}, {Buchhave}, {Cameron},
  {Carleo}, {Charbonneau}, {Claudi}, {Cosentino}, {Covino}, {Delfosse},
  {Desidera}, {Di Fabrizio}, {Dressing}, {Esposito}, {Fares}, {Figueira},
  {Fiorenzano}, {Forveille}, {Giacobbe}, {Gonz{\'a}lez-{\'A}lvarez}, {Gratton},
  {Harutyunyan}, {Johnson}, {Latham}, {Leto}, {Lopez-Morales}, {Lovis},
  {Maggio}, {Mancini}, {Masiero}, {Mayor}, {Micela}, {Molinari}, {Motalebi},
  {Murgas}, {Nascimbeni}, {Pagano}, {Pepe}, {Phillips}, {Piotto}, {Poretti},
  {Rainer}, {Rice}, {Santos}, {Sasselov}, {Scandariato}, {S{\'e}gransan},
  {Smareglia}, {Udry}, {Watson}, \& {W{\"u}nsche}}]{Damasso2018}
{Damasso}, M., {Bonomo}, A.~S., {Astudillo-Defru}, N., {et~al.} 2018, \aap,
  615, A69, \dodoi{10.1051/0004-6361/201732459}

\bibitem[{{Del Zanna} {et~al.}(2021){Del Zanna}, {Dere}, {Young}, \&
  {Landi}}]{DelZanna2021}
{Del Zanna}, G., {Dere}, K.~P., {Young}, P.~R., \& {Landi}, E. 2021, \apj, 909,
  38, \dodoi{10.3847/1538-4357/abd8ce}

\bibitem[{{Dere} {et~al.}(1997){Dere}, {Landi}, {Mason}, {Monsignori Fossi}, \&
  {Young}}]{Dere1997}
{Dere}, K.~P., {Landi}, E., {Mason}, H.~E., {Monsignori Fossi}, B.~C., \&
  {Young}, P.~R. 1997, \aaps, 125, 149, \dodoi{10.1051/aas:1997368}

\bibitem[{Diamond-Lowe {et~al.}(2021)Diamond-Lowe, Youngblood, Charbonneau,
  King, Teal, Bastelberger, Corrales, \& Kempton}]{Diamond-Lowe2021}
Diamond-Lowe, H., Youngblood, A., Charbonneau, D., {et~al.} 2021, The
  Astronomical Journal, 162, 10, \dodoi{10.3847/1538-3881/abfa1c}

\bibitem[{{Dorn} \& {Lichtenberg}(2021)}]{Dorn2021}
{Dorn}, C., \& {Lichtenberg}, T. 2021, \apjl, 922, L4,
  \dodoi{10.3847/2041-8213/ac33af}

\bibitem[{{Dotter}(2016)}]{Dotter2016}
{Dotter}, A. 2016, \apjs, 222, 8, \dodoi{10.3847/0067-0049/222/1/8}

\bibitem[{{Douglas} {et~al.}(2014){Douglas}, {Ag{\"u}eros}, {Covey}, {Bowsher},
  {Bochanski}, {Cargile}, {Kraus}, {Law}, {Lemonias}, {Arce}, {Fierroz}, \&
  {Kundert}}]{Douglas2014}
{Douglas}, S.~T., {Ag{\"u}eros}, M.~A., {Covey}, K.~R., {et~al.} 2014, \apj,
  795, 161, \dodoi{10.1088/0004-637X/795/2/161}

\bibitem[{{Duvvuri} {et~al.}(2021){Duvvuri}, {Sebastian Pineda},
  {Berta-Thompson}, {Brown}, {France}, {Kowalski}, {Redfield}, {Tilipman},
  {Vieytes}, {Wilson}, {Youngblood}, {Froning}, {Linsky}, {Parke Loyd},
  {Mauas}, {Miguel}, {Newton}, {Rugheimer}, \& {Christian
  Schneider}}]{Duvvuri2021}
{Duvvuri}, G.~M., {Sebastian Pineda}, J., {Berta-Thompson}, Z.~K., {et~al.}
  2021, \apj, 913, 40, \dodoi{10.3847/1538-4357/abeaaf}

\bibitem[{{Eastman}(2017)}]{Eastman2017}
{Eastman}, J. 2017, {EXOFASTv2: Generalized publication-quality exoplanet
  modeling code}.
\newblock \doeprint{1710.003}

\bibitem[{{Eastman} {et~al.}(2013){Eastman}, {Gaudi}, \& {Agol}}]{Eastman2013}
{Eastman}, J., {Gaudi}, B.~S., \& {Agol}, E. 2013, \pasp, 125, 83,
  \dodoi{10.1086/669497}

\bibitem[{{Eastman} {et~al.}(2019){Eastman}, {Rodriguez}, {Agol}, {Stassun},
  {Beatty}, {Vanderburg}, {Gaudi}, {Collins}, \& {Luger}}]{Eastman2019}
{Eastman}, J.~D., {Rodriguez}, J.~E., {Agol}, E., {et~al.} 2019, arXiv
  e-prints, arXiv:1907.09480.
\newblock \doarXiv{1907.09480}

\bibitem[{{Erkaev} {et~al.}(2007){Erkaev}, {Kulikov}, {Lammer}, {Selsis},
  {Langmayr}, {Jaritz}, \& {Biernat}}]{Erkaev2007}
{Erkaev}, N.~V., {Kulikov}, Y.~N., {Lammer}, H., {et~al.} 2007, \aap, 472, 329,
  \dodoi{10.1051/0004-6361:20066929}

\bibitem[{{France} {et~al.}(2016){France}, {Loyd}, {Youngblood}, {Brown},
  {Schneider}, {Hawley}, {Froning}, {Linsky}, {Roberge}, {Buccino},
  {Davenport}, {Fontenla}, {Kaltenegger}, {Kowalski}, {Mauas}, {Miguel},
  {Redfield}, {Rugheimer}, {Tian}, {Vieytes}, {Walkowicz}, \&
  {Weisenburger}}]{France2016}
{France}, K., {Loyd}, R.~O.~P., {Youngblood}, A., {et~al.} 2016, \apj, 820, 89,
  \dodoi{10.3847/0004-637X/820/2/89}

\bibitem[{{Fulton} \& {Petigura}(2018)}]{Fulton2018}
{Fulton}, B.~J., \& {Petigura}, E.~A. 2018, \aj, 156, 264,
  \dodoi{10.3847/1538-3881/aae828}

\bibitem[{{Fulton} {et~al.}(2017){Fulton}, {Petigura}, {Howard}, {Isaacson},
  {Marcy}, {Cargile}, {Hebb}, {Weiss}, {Johnson}, {Morton}, {Sinukoff},
  {Crossfield}, \& {Hirsch}}]{Fulton2017}
{Fulton}, B.~J., {Petigura}, E.~A., {Howard}, A.~W., {et~al.} 2017, \aj, 154,
  109, \dodoi{10.3847/1538-3881/aa80eb}

\bibitem[{{Gaia Collaboration} {et~al.}(2016){Gaia Collaboration}, {Prusti},
  {de Bruijne}, {Brown}, {Vallenari}, {Babusiaux}, {Bailer-Jones}, {Bastian},
  {Biermann}, {Evans}, {Eyer}, {Jansen}, {Jordi}, {Klioner}, {Lammers},
  {Lindegren}, {Luri}, {Mignard}, {Milligan}, {Panem}, {Poinsignon},
  {Pourbaix}, {Randich}, {Sarri}, {Sartoretti}, {Siddiqui}, {Soubiran},
  {Valette}, {van Leeuwen}, {Walton}, {Aerts}, {Arenou}, {Cropper}, {Drimmel},
  {H{\o}g}, {Katz}, {Lattanzi}, {O'Mullane}, {Grebel}, {Holland}, {Huc},
  {Passot}, {Bramante}, {Cacciari}, {Casta{\~n}eda}, {Chaoul}, {Cheek}, {De
  Angeli}, {Fabricius}, {Guerra}, {Hern{\'a}ndez}, {Jean-Antoine-Piccolo},
  {Masana}, {Messineo}, {Mowlavi}, {Nienartowicz}, {Ord{\'o}{\~n}ez-Blanco},
  {Panuzzo}, {Portell}, {Richards}, {Riello}, {Seabroke}, {Tanga},
  {Th{\'e}venin}, {Torra}, {Els}, {Gracia-Abril}, {Comoretto},
  {Garcia-Reinaldos}, {Lock}, {Mercier}, {Altmann}, {Andrae}, {Astraatmadja},
  {Bellas-Velidis}, {Benson}, {Berthier}, {Blomme}, {Busso}, {Carry},
  {Cellino}, {Clementini}, {Cowell}, {Creevey}, {Cuypers}, {Davidson}, {De
  Ridder}, {de Torres}, {Delchambre}, {Dell'Oro}, {Ducourant}, {Fr{\'e}mat},
  {Garc{\'\i}a-Torres}, {Gosset}, {Halbwachs}, {Hambly}, {Harrison}, {Hauser},
  {Hestroffer}, {Hodgkin}, {Huckle}, {Hutton}, {Jasniewicz}, {Jordan},
  {Kontizas}, {Korn}, {Lanzafame}, {Manteiga}, {Moitinho}, {Muinonen},
  {Osinde}, {Pancino}, {Pauwels}, {Petit}, {Recio-Blanco}, {Robin}, {Sarro},
  {Siopis}, {Smith}, {Smith}, {Sozzetti}, {Thuillot}, {van Reeven}, {Viala},
  {Abbas}, {Abreu Aramburu}, {Accart}, {Aguado}, {Allan}, {Allasia},
  {Altavilla}, {{\'A}lvarez}, {Alves}, {Anderson}, {Andrei}, {Anglada Varela},
  {Antiche}, {Antoja}, {Ant{\'o}n}, {Arcay}, {Atzei}, {Ayache}, {Bach},
  {Baker}, {Balaguer-N{\'u}{\~n}ez}, {Barache}, {Barata}, {Barbier}, {Barblan},
  {Baroni}, {Barrado y Navascu{\'e}s}, {Barros}, {Barstow}, {Becciani},
  {Bellazzini}, {Bellei}, {Bello Garc{\'\i}a}, {Belokurov}, {Bendjoya},
  {Berihuete}, {Bianchi}, {Bienaym{\'e}}, {Billebaud}, {Blagorodnova},
  {Blanco-Cuaresma}, {Boch}, {Bombrun}, {Borrachero}, {Bouquillon}, {Bourda},
  {Bouy}, {Bragaglia}, {Breddels}, {Brouillet}, {Br{\"u}semeister},
  {Bucciarelli}, {Budnik}, {Burgess}, {Burgon}, {Burlacu}, {Busonero}, {Buzzi},
  {Caffau}, {Cambras}, {Campbell}, {Cancelliere}, {Cantat-Gaudin}, {Carlucci},
  {Carrasco}, {Castellani}, {Charlot}, {Charnas}, {Charvet}, {Chassat},
  {Chiavassa}, {Clotet}, {Cocozza}, {Collins}, {Collins}, {Costigan}, {Crifo},
  {Cross}, {Crosta}, {Crowley}, {Dafonte}, {Damerdji}, {Dapergolas}, {David},
  {David}, {De Cat}, {de Felice}, {de Laverny}, {De Luise}, {De March}, {de
  Martino}, {de Souza}, {Debosscher}, {del Pozo}, {Delbo}, {Delgado},
  {Delgado}, {di Marco}, {Di Matteo}, {Diakite}, {Distefano}, {Dolding}, {Dos
  Anjos}, {Drazinos}, {Dur{\'a}n}, {Dzigan}, {Ecale}, {Edvardsson}, {Enke},
  {Erdmann}, {Escolar}, {Espina}, {Evans}, {Eynard Bontemps}, {Fabre},
  {Fabrizio}, {Faigler}, {Falc{\~a}o}, {Farr{\`a}s Casas}, {Faye}, {Federici},
  {Fedorets}, {Fern{\'a}ndez-Hern{\'a}ndez}, {Fernique}, {Fienga}, {Figueras},
  {Filippi}, {Findeisen}, {Fonti}, {Fouesneau}, {Fraile}, {Fraser}, {Fuchs},
  {Furnell}, {Gai}, {Galleti}, {Galluccio}, {Garabato}, {Garc{\'\i}a-Sedano},
  {Gar{\'e}}, {Garofalo}, {Garralda}, {Gavras}, {Gerssen}, {Geyer}, {Gilmore},
  {Girona}, {Giuffrida}, {Gomes}, {Gonz{\'a}lez-Marcos},
  {Gonz{\'a}lez-N{\'u}{\~n}ez}, {Gonz{\'a}lez-Vidal}, {Granvik}, {Guerrier},
  {Guillout}, {Guiraud}, {G{\'u}rpide}, {Guti{\'e}rrez-S{\'a}nchez}, {Guy},
  {Haigron}, {Hatzidimitriou}, {Haywood}, {Heiter}, {Helmi}, {Hobbs},
  {Hofmann}, {Holl}, {Holland }, {Hunt}, {Hypki}, {Icardi}, {Irwin}, {Jevardat
  de Fombelle}, {Jofr{\'e}}, {Jonker}, {Jorissen}, {Julbe}, {Karampelas},
  {Kochoska}, {Kohley}, {Kolenberg}, {Kontizas}, {Koposov}, {Kordopatis},
  {Koubsky}, {Kowalczyk}, {Krone-Martins}, {Kudryashova}, {Kull}, {Bachchan},
  {Lacoste-Seris}, {Lanza}, {Lavigne}, {Le Poncin-Lafitte}, {Lebreton},
  {Lebzelter}, {Leccia}, {Leclerc}, {Lecoeur-Taibi}, {Lemaitre}, {Lenhardt},
  {Leroux}, {Liao}, {Licata}, {Lindstr{\o}m}, {Lister}, {Livanou}, {Lobel},
  {L{\"o}ffler}, {L{\'o}pez}, {Lopez-Lozano}, {Lorenz}, {Loureiro},
  {MacDonald}, {Magalh{\~a}es Fernandes}, {Managau}, {Mann}, {Mantelet},
  {Marchal}, {Marchant}, {Marconi}, {Marie}, {Marinoni}, {Marrese},
  {Marschalk{\'o}}, {Marshall}, {Mart{\'\i}n-Fleitas}, {Martino}, {Mary},
  {Matijevi{\v{c}}}, {Mazeh}, {McMillan}, {Messina}, {Mestre}, {Michalik},
  {Millar}, {Miranda}, {Molina}, {Molinaro}, {Molinaro}, {Moln{\'a}r},
  {Moniez}, {Montegriffo}, {Monteiro}, {Mor}, {Mora}, {Morbidelli}, {Morel},
  {Morgenthaler}, {Morley}, {Morris}, {Mulone}, {Muraveva}, {Musella},
  {Narbonne}, {Nelemans}, {Nicastro}, {Noval}, {Ord{\'e}novic},
  {Ordieres-Mer{\'e}}, {Osborne}, {Pagani}, {Pagano}, {Pailler}, {Palacin},
  {Palaversa}, {Parsons}, {Paulsen}, {Pecoraro}, {Pedrosa}, {Pentik{\"a}inen},
  {Pereira}, {Pichon}, {Piersimoni}, {Pineau}, {Plachy}, {Plum}, {Poujoulet},
  {Pr{\v{s}}a}, {Pulone}, {Ragaini}, {Rago}, {Rambaux}, {Ramos-Lerate},
  {Ranalli}, {Rauw}, {Read}, {Regibo}, {Renk}, {Reyl{\'e}}, {Ribeiro},
  {Rimoldini}, {Ripepi}, {Riva}, {Rixon}, {Roelens}, {Romero-G{\'o}mez},
  {Rowell}, {Royer}, {Rudolph}, {Ruiz-Dern}, {Sadowski}, {Sagrist{\`a}
  Sell{\'e}s}, {Sahlmann}, {Salgado}, {Salguero}, {Sarasso}, {Savietto},
  {Schnorhk}, {Schultheis}, {Sciacca}, {Segol}, {Segovia}, {Segransan},
  {Serpell}, {Shih}, {Smareglia}, {Smart}, {Smith}, {Solano}, {Solitro},
  {Sordo}, {Soria Nieto}, {Souchay}, {Spagna}, {Spoto}, {Stampa}, {Steele},
  {Steidelm{\"u}ller}, {Stephenson}, {Stoev}, {Suess}, {S{\"u}veges}, {Surdej},
  {Szabados}, {Szegedi-Elek}, {Tapiador}, {Taris}, {Tauran}, {Taylor},
  {Teixeira}, {Terrett}, {Tingley}, {Trager}, {Turon}, {Ulla}, {Utrilla},
  {Valentini}, {van Elteren}, {Van Hemelryck}, {van Leeuwen}, {Varadi},
  {Vecchiato}, {Veljanoski}, {Via}, {Vicente}, {Vogt}, {Voss}, {Votruba},
  {Voutsinas}, {Walmsley}, {Weiler}, {Weingrill}, {Werner}, {Wevers},
  {Whitehead}, {Wyrzykowski}, {Yoldas}, {{\v{Z}}erjal}, {Zucker}, {Zurbach},
  {Zwitter}, {Alecu}, {Allen}, {Allende Prieto}, {Amorim},
  {Anglada-Escud{\'e}}, {Arsenijevic}, {Azaz}, {Balm}, {Beck}, {Bernstein},
  {Bigot}, {Bijaoui}, {Blasco}, {Bonfigli}, {Bono}, {Boudreault}, {Bressan},
  {Brown}, {Brunet}, {Bunclark}, {Buonanno}, {Butkevich}, {Carret}, {Carrion},
  {Chemin}, {Ch{\'e}reau}, {Corcione}, {Darmigny}, {de Boer}, {de Teodoro}, {de
  Zeeuw}, {Delle Luche}, {Domingues}, {Dubath}, {Fodor}, {Fr{\'e}zouls},
  {Fries}, {Fustes}, {Fyfe}, {Gallardo}, {Gallegos}, {Gardiol}, {Gebran},
  {Gomboc}, {G{\'o}mez}, {Grux}, {Gueguen}, {Heyrovsky}, {Hoar}, {Iannicola},
  {Isasi Parache}, {Janotto}, {Joliet}, {Jonckheere}, {Keil}, {Kim},
  {Klagyivik}, {Klar}, {Knude}, {Kochukhov}, {Kolka}, {Kos}, {Kutka}, {Lainey},
  {LeBouquin}, {Liu}, {Loreggia}, {Makarov}, {Marseille}, {Martayan},
  {Martinez-Rubi}, {Massart}, {Meynadier}, {Mignot}, {Munari}, {Nguyen},
  {Nordlander}, {Ocvirk}, {O'Flaherty}, {Olias Sanz}, {Ortiz}, {Osorio},
  {Oszkiewicz}, {Ouzounis}, {Palmer}, {Park}, {Pasquato}, {Peltzer}, {Peralta},
  {P{\'e}turaud}, {Pieniluoma}, {Pigozzi}, {Poels}, {Prat}, {Prod'homme},
  {Raison}, {Rebordao}, {Risquez}, {Rocca-Volmerange}, {Rosen}, {Ruiz-Fuertes},
  {Russo}, {Sembay}, {Serraller Vizcaino}, {Short}, {Siebert}, {Silva},
  {Sinachopoulos}, {Slezak}, {Soffel}, {Sosnowska}, {Strai{\v{z}}ys}, {ter
  Linden}, {Terrell}, {Theil}, {Tiede}, {Troisi}, {Tsalmantza}, {Tur},
  {Vaccari}, {Vachier}, {Valles}, {Van Hamme}, {Veltz}, {Virtanen}, {Wallut},
  {Wichmann}, {Wilkinson}, {Ziaeepour}, \& {Zschocke}}]{GaiaMission2016}
{Gaia Collaboration}, {Prusti}, T., {de Bruijne}, J.~H.~J., {et~al.} 2016,
  \aap, 595, A1, \dodoi{10.1051/0004-6361/201629272}

\bibitem[{{Gaia Collaboration} {et~al.}(2018){Gaia Collaboration}, {Brown},
  {Vallenari}, {Prusti}, {de Bruijne}, {Babusiaux}, {Bailer-Jones}, {Biermann},
  {Evans}, {Eyer}, {Jansen}, {Jordi}, {Klioner}, {Lammers}, {Lindegren},
  {Luri}, {Mignard}, {Panem}, {Pourbaix}, {Randich}, {Sartoretti}, {Siddiqui},
  {Soubiran}, {van Leeuwen}, {Walton}, {Arenou}, {Bastian}, {Cropper},
  {Drimmel}, {Katz}, {Lattanzi}, {Bakker}, {Cacciari}, {Casta{\~n}eda},
  {Chaoul}, {Cheek}, {De Angeli}, {Fabricius}, {Guerra}, {Holl}, {Masana},
  {Messineo}, {Mowlavi}, {Nienartowicz}, {Panuzzo}, {Portell}, {Riello},
  {Seabroke}, {Tanga}, {Th{\'e}venin}, {Gracia-Abril}, {Comoretto},
  {Garcia-Reinaldos}, {Teyssier}, {Altmann}, {Andrae}, {Audard},
  {Bellas-Velidis}, {Benson}, {Berthier}, {Blomme}, {Burgess}, {Busso},
  {Carry}, {Cellino}, {Clementini}, {Clotet}, {Creevey}, {Davidson}, {De
  Ridder}, {Delchambre}, {Dell'Oro}, {Ducourant},
  {Fern{\'a}ndez-Hern{\'a}ndez}, {Fouesneau}, {Fr{\'e}mat}, {Galluccio},
  {Garc{\'\i}a-Torres}, {Gonz{\'a}lez-N{\'u}{\~n}ez}, {Gonz{\'a}lez-Vidal},
  {Gosset}, {Guy}, {Halbwachs}, {Hambly}, {Harrison}, {Hern{\'a}ndez},
  {Hestroffer}, {Hodgkin}, {Hutton}, {Jasniewicz}, {Jean-Antoine-Piccolo},
  {Jordan}, {Korn}, {Krone-Martins}, {Lanzafame}, {Lebzelter}, {L{\"o}ffler},
  {Manteiga}, {Marrese}, {Mart{\'\i}n-Fleitas}, {Moitinho}, {Mora}, {Muinonen},
  {Osinde}, {Pancino}, {Pauwels}, {Petit}, {Recio-Blanco}, {Richards},
  {Rimoldini}, {Robin}, {Sarro}, {Siopis}, {Smith}, {Sozzetti}, {S{\"u}veges},
  {Torra}, {van Reeven}, {Abbas}, {Abreu Aramburu}, {Accart}, {Aerts},
  {Altavilla}, {{\'A}lvarez}, {Alvarez}, {Alves}, {Anderson}, {Andrei},
  {Anglada Varela}, {Antiche}, {Antoja}, {Arcay}, {Astraatmadja}, {Bach},
  {Baker}, {Balaguer-N{\'u}{\~n}ez}, {Balm}, {Barache}, {Barata}, {Barbato},
  {Barblan}, {Barklem}, {Barrado}, {Barros}, {Barstow}, {Bartholom{\'e}
  Mu{\~n}oz}, {Bassilana}, {Becciani}, {Bellazzini}, {Berihuete}, {Bertone},
  {Bianchi}, {Bienaym{\'e}}, {Blanco-Cuaresma}, {Boch}, {Boeche}, {Bombrun},
  {Borrachero}, {Bossini}, {Bouquillon}, {Bourda}, {Bragaglia}, {Bramante},
  {Breddels}, {Bressan}, {Brouillet}, {Br{\"u}semeister}, {Brugaletta},
  {Bucciarelli}, {Burlacu}, {Busonero}, {Butkevich}, {Buzzi}, {Caffau},
  {Cancelliere}, {Cannizzaro}, {Cantat-Gaudin}, {Carballo}, {Carlucci},
  {Carrasco}, {Casamiquela}, {Castellani}, {Castro-Ginard}, {Charlot},
  {Chemin}, {Chiavassa}, {Cocozza}, {Costigan}, {Cowell}, {Crifo}, {Crosta},
  {Crowley}, {Cuypers}, {Dafonte}, {Damerdji}, {Dapergolas}, {David}, {David},
  {de Laverny}, {De Luise}, {De March}, {de Martino}, {de Souza}, {de Torres},
  {Debosscher}, {del Pozo}, {Delbo}, {Delgado}, {Delgado}, {Di Matteo},
  {Diakite}, {Diener}, {Distefano}, {Dolding}, {Drazinos}, {Dur{\'a}n},
  {Edvardsson}, {Enke}, {Eriksson}, {Esquej}, {Eynard Bontemps}, {Fabre},
  {Fabrizio}, {Faigler}, {Falc{\~a}o}, {Farr{\`a}s Casas}, {Federici},
  {Fedorets}, {Fernique}, {Figueras}, {Filippi}, {Findeisen}, {Fonti},
  {Fraile}, {Fraser}, {Fr{\'e}zouls}, {Gai}, {Galleti}, {Garabato},
  {Garc{\'\i}a-Sedano}, {Garofalo}, {Garralda}, {Gavel}, {Gavras}, {Gerssen},
  {Geyer}, {Giacobbe}, {Gilmore}, {Girona}, {Giuffrida}, {Glass}, {Gomes},
  {Granvik}, {Gueguen}, {Guerrier}, {Guiraud}, {Guti{\'e}rrez-S{\'a}nchez},
  {Haigron}, {Hatzidimitriou}, {Hauser}, {Haywood}, {Heiter}, {Helmi}, {Heu},
  {Hilger}, {Hobbs}, {Hofmann}, {Holland}, {Huckle}, {Hypki}, {Icardi},
  {Jan{\ss}en}, {Jevardat de Fombelle}, {Jonker}, {Juh{\'a}sz}, {Julbe},
  {Karampelas}, {Kewley}, {Klar}, {Kochoska}, {Kohley}, {Kolenberg},
  {Kontizas}, {Kontizas}, {Koposov}, {Kordopatis}, {Kostrzewa-Rutkowska},
  {Koubsky}, {Lambert}, {Lanza}, {Lasne}, {Lavigne}, {Le Fustec}, {Le
  Poncin-Lafitte}, {Lebreton}, {Leccia}, {Leclerc}, {Lecoeur-Taibi},
  {Lenhardt}, {Leroux}, {Liao}, {Licata}, {Lindstr{\o}m}, {Lister}, {Livanou},
  {Lobel}, {L{\'o}pez}, {Managau}, {Mann}, {Mantelet}, {Marchal}, {Marchant},
  {Marconi}, {Marinoni}, {Marschalk{\'o}}, {Marshall}, {Martino}, {Marton},
  {Mary}, {Massari}, {Matijevi{\v{c}}}, {Mazeh}, {McMillan}, {Messina},
  {Michalik}, {Millar}, {Molina}, {Molinaro}, {Moln{\'a}r}, {Montegriffo},
  {Mor}, {Morbidelli}, {Morel}, {Morris}, {Mulone}, {Muraveva}, {Musella},
  {Nelemans}, {Nicastro}, {Noval}, {O'Mullane}, {Ord{\'e}novic},
  {Ord{\'o}{\~n}ez-Blanco}, {Osborne}, {Pagani}, {Pagano}, {Pailler},
  {Palacin}, {Palaversa}, {Panahi}, {Pawlak}, {Piersimoni}, {Pineau}, {Plachy},
  {Plum}, {Poggio}, {Poujoulet}, {Pr{\v{s}}a}, {Pulone}, {Racero}, {Ragaini},
  {Rambaux}, {Ramos-Lerate}, {Regibo}, {Reyl{\'e}}, {Riclet}, {Ripepi}, {Riva},
  {Rivard}, {Rixon}, {Roegiers}, {Roelens}, {Romero-G{\'o}mez}, {Rowell},
  {Royer}, {Ruiz-Dern}, {Sadowski}, {Sagrist{\`a} Sell{\'e}s}, {Sahlmann},
  {Salgado}, {Salguero}, {Sanna}, {Santana-Ros}, {Sarasso}, {Savietto},
  {Schultheis}, {Sciacca}, {Segol}, {Segovia}, {S{\'e}gransan}, {Shih},
  {Siltala}, {Silva}, {Smart}, {Smith}, {Solano}, {Solitro}, {Sordo}, {Soria
  Nieto}, {Souchay}, {Spagna}, {Spoto}, {Stampa}, {Steele},
  {Steidelm{\"u}ller}, {Stephenson}, {Stoev}, {Suess}, {Surdej}, {Szabados},
  {Szegedi-Elek}, {Tapiador}, {Taris}, {Tauran}, {Taylor}, {Teixeira},
  {Terrett}, {Teyssand ier}, {Thuillot}, {Titarenko}, {Torra Clotet}, {Turon},
  {Ulla}, {Utrilla}, {Uzzi}, {Vaillant}, {Valentini}, {Valette}, {van Elteren},
  {Van Hemelryck}, {van Leeuwen}, {Vaschetto}, {Vecchiato}, {Veljanoski},
  {Viala}, {Vicente}, {Vogt}, {von Essen}, {Voss}, {Votruba}, {Voutsinas},
  {Walmsley}, {Weiler}, {Wertz}, {Wevers}, {Wyrzykowski}, {Yoldas},
  {{\v{Z}}erjal}, {Ziaeepour}, {Zorec}, {Zschocke}, {Zucker}, {Zurbach}, \&
  {Zwitter}}]{GaiaDR22018}
{Gaia Collaboration}, {Brown}, A.~G.~A., {Vallenari}, A., {et~al.} 2018, \aap,
  616, A1, \dodoi{10.1051/0004-6361/201833051}

\bibitem[{{Ginzburg} {et~al.}(2016){Ginzburg}, {Schlichting}, \&
  {Sari}}]{Ginzburg2016}
{Ginzburg}, S., {Schlichting}, H.~E., \& {Sari}, R. 2016, \apj, 825, 29,
  \dodoi{10.3847/0004-637X/825/1/29}

\bibitem[{{Ginzburg} {et~al.}(2018){Ginzburg}, {Schlichting}, \&
  {Sari}}]{Ginzburg2018}
---. 2018, \mnras, 476, 759, \dodoi{10.1093/mnras/sty290}

\bibitem[{{Green} {et~al.}(2012){Green}, {Froning}, {Osterman}, {Ebbets},
  {Heap}, {Leitherer}, {Linsky}, {Savage}, {Sembach}, {Shull}, {Siegmund},
  {Snow}, {Spencer}, {Stern}, {Stocke}, {Welsh}, {B{\'e}land}, {Burgh},
  {Danforth}, {France}, {Keeney}, {McPhate}, {Penton}, {Andrews},
  {Brownsberger}, {Morse}, \& {Wilkinson}}]{Green2012}
{Green}, J.~C., {Froning}, C.~S., {Osterman}, S., {et~al.} 2012, \apj, 744, 60,
  \dodoi{10.1088/0004-637X/744/1/60}

\bibitem[{{Gupta} \& {Schlichting}(2019)}]{Gupta&Schlichting2019}
{Gupta}, A., \& {Schlichting}, H.~E. 2019, \mnras, 487, 24,
  \dodoi{10.1093/mnras/stz1230}

\bibitem[{{Hansen} \& {Barman}(2007)}]{Hansen2007}
{Hansen}, B. M.~S., \& {Barman}, T. 2007, \apj, 671, 861,
  \dodoi{10.1086/523038}

\bibitem[{Harman {et~al.}(2022)Harman, Kopparapu, Stef{\'{a}}nsson, Lin,
  Mahadevan, Hedges, \& Batalha}]{Harman2022}
Harman, C.~E., Kopparapu, R.~K., Stef{\'{a}}nsson, G., {et~al.} 2022, \psj, 3,
  45, \dodoi{10.3847/psj/ac38ac}

\bibitem[{{Howell} {et~al.}(2014){Howell}, {Sobeck}, {Haas}, {Still},
  {Barclay}, {Mullally}, {Troeltzsch}, {Aigrain}, {Bryson}, {Caldwell},
  {Chaplin}, {Cochran}, {Huber}, {Marcy}, {Miglio}, {Najita}, {Smith},
  {Twicken}, \& {Fortney}}]{Howell2014}
{Howell}, S.~B., {Sobeck}, C., {Haas}, M., {et~al.} 2014, \pasp, 126, 398,
  \dodoi{10.1086/676406}

\bibitem[{{Kanodia} {et~al.}(2019){Kanodia}, {Wolfgang}, {Stefansson}, {Ning},
  \& {Mahadevan}}]{Kanodia2019}
{Kanodia}, S., {Wolfgang}, A., {Stefansson}, G.~K., {Ning}, B., \& {Mahadevan},
  S. 2019, \apj, 882, 38, \dodoi{10.3847/1538-4357/ab334c}

\bibitem[{{Kasting}(1988)}]{Kasting1988}
{Kasting}, J.~F. 1988, \icarus, 74, 472, \dodoi{10.1016/0019-1035(88)90116-9}

\bibitem[{{Kasting} \& {Pollack}(1983)}]{Kasting1983}
{Kasting}, J.~F., \& {Pollack}, J.~B. 1983, \icarus, 53, 479,
  \dodoi{10.1016/0019-1035(83)90212-9}

\bibitem[{{Kopparapu} {et~al.}(2014){Kopparapu}, {Ramirez}, {SchottelKotte},
  {Kasting}, {Domagal-Goldman}, \& {Eymet}}]{Kopparapu2014}
{Kopparapu}, R.~K., {Ramirez}, R.~M., {SchottelKotte}, J., {et~al.} 2014,
  \apjl, 787, L29, \dodoi{10.1088/2041-8205/787/2/L29}

\bibitem[{Kopparapu {et~al.}(2017)Kopparapu, Wolf, Arney, Batalha, Haqq-Misra,
  Grimm, \& Heng}]{Kopparapu2017}
Kopparapu, R.~K., Wolf, E.~T., Arney, G., {et~al.} 2017, \apj, 845, 5,
  \dodoi{10.3847/1538-4357/aa7cf9}

\bibitem[{{Kosiarek} {et~al.}(2019){Kosiarek}, {Crossfield},
  {Hardegree-Ullman}, {Livingston}, {Benneke}, {Henry}, {Howard}, {Berardo},
  {Blunt}, {Fulton}, {Hirsch}, {Howard}, {Isaacson}, {Petigura}, {Sinukoff},
  {Weiss}, {Bonfils}, {Dressing}, {Knutson}, {Schlieder}, {Werner}, {Gorjian},
  {Krick}, {Morales}, {Astudillo-Defru}, {Almenara}, {Delfosse}, {Forveille},
  {Lovis}, {Mayor}, {Murgas}, {Pepe}, {Santos}, {Udry}, {Corbett}, {Fors},
  {Law}, {Ratzloff}, \& {del Ser}}]{Kosiarek2019}
{Kosiarek}, M.~R., {Crossfield}, I. J.~M., {Hardegree-Ullman}, K.~K., {et~al.}
  2019, \aj, 157, 97, \dodoi{10.3847/1538-3881/aaf79c}

\bibitem[{{Landi} {et~al.}(2012){Landi}, {Del Zanna}, {Young}, {Dere}, \&
  {Mason}}]{Landi2012}
{Landi}, E., {Del Zanna}, G., {Young}, P.~R., {Dere}, K.~P., \& {Mason}, H.~E.
  2012, \apj, 744, 99, \dodoi{10.1088/0004-637X/744/2/99}

\bibitem[{{Lee} \& {Chiang}(2016)}]{Lee&Chiang2016}
{Lee}, E.~J., \& {Chiang}, E. 2016, \apj, 817, 90,
  \dodoi{10.3847/0004-637X/817/2/90}

\bibitem[{{Lee} \& {Connors}(2021)}]{Lee2021}
{Lee}, E.~J., \& {Connors}, N.~J. 2021, \apj, 908, 32,
  \dodoi{10.3847/1538-4357/abd6c7}

\bibitem[{{Lee} {et~al.}(2022){Lee}, {Karalis}, \& {Thorngren}}]{Lee2022}
{Lee}, E.~J., {Karalis}, A., \& {Thorngren}, D.~P. 2022, arXiv e-prints,
  arXiv:2201.09898.
\newblock \doarXiv{2201.09898}

\bibitem[{{Lopez} \& {Fortney}(2013)}]{Lopez2013}
{Lopez}, E.~D., \& {Fortney}, J.~J. 2013, \apj, 776, 2,
  \dodoi{10.1088/0004-637X/776/1/2}

\bibitem[{{Lopez} \& {Rice}(2018)}]{Lopez2018}
{Lopez}, E.~D., \& {Rice}, K. 2018, \mnras, 479, 5303,
  \dodoi{10.1093/mnras/sty1707}

\bibitem[{{Louden} {et~al.}(2017){Louden}, {Wheatley}, \&
  {Briggs}}]{Louden2017}
{Louden}, T., {Wheatley}, P.~J., \& {Briggs}, K. 2017, \mnras, 464, 2396,
  \dodoi{10.1093/mnras/stw2421}

\bibitem[{{Loyd} {et~al.}(2016){Loyd}, {France}, {Youngblood}, {Schneider},
  {Brown}, {Hu}, {Linsky}, {Froning}, {Redfield}, {Rugheimer}, \&
  {Tian}}]{Loyd2016}
{Loyd}, R.~O.~P., {France}, K., {Youngblood}, A., {et~al.} 2016, \apj, 824,
  102, \dodoi{10.3847/0004-637X/824/2/102}

\bibitem[{{Malik} {et~al.}(2017){Malik}, {Grosheintz}, {Mendon{\c{c}}a},
  {Grimm}, {Lavie}, {Kitzmann}, {Tsai}, {Burrows}, {Kreidberg}, {Bedell},
  {Bean}, {Stevenson}, \& {Heng}}]{Malik2017}
{Malik}, M., {Grosheintz}, L., {Mendon{\c{c}}a}, J.~M., {et~al.} 2017, \aj,
  153, 56, \dodoi{10.3847/1538-3881/153/2/56}

\bibitem[{{Mann} {et~al.}(2019){Mann}, {Dupuy}, {Kraus}, {Gaidos}, {Ansdell},
  {Ireland}, {Rizzuto}, {Hung}, {Dittmann}, {Factor}, {Feiden}, {Martinez},
  {Ru{\'\i}z-Rodr{\'\i}guez}, \& {Thao}}]{Mann2019}
{Mann}, A.~W., {Dupuy}, T., {Kraus}, A.~L., {et~al.} 2019, \apj, 871, 63,
  \dodoi{10.3847/1538-4357/aaf3bc}

\bibitem[{{Martinez} {et~al.}(2019){Martinez}, {Cunha}, {Ghezzi}, \&
  {Smith}}]{Martinez2019}
{Martinez}, C.~F., {Cunha}, K., {Ghezzi}, L., \& {Smith}, V.~V. 2019, \apj,
  875, 29, \dodoi{10.3847/1538-4357/ab0d93}

\bibitem[{{Melbourne} {et~al.}(2020){Melbourne}, {Youngblood}, {France},
  {Froning}, {Pineda}, {Shkolnik}, {Wilson}, {Wood}, {Basu}, {Roberge},
  {Schlieder}, {Cauley}, {Loyd}, {Newton}, {Schneider}, {Arulanantham},
  {Berta-Thompson}, {Brown}, {Buccino}, {Kempton}, {Linsky}, {Logsdon},
  {Mauas}, {Pagano}, {Peacock}, {Redfield}, {Rugheimer}, {Schneider}, {Teal},
  {Tian}, {Tilipman}, \& {Vieytes}}]{Melbourne2020}
{Melbourne}, K., {Youngblood}, A., {France}, K., {et~al.} 2020, \aj, 160, 269,
  \dodoi{10.3847/1538-3881/abbf5c}

\bibitem[{{Millholland} \& {Winn}(2021)}]{Millholland2021}
{Millholland}, S.~C., \& {Winn}, J.~N. 2021, \apjl, 920, L34,
  \dodoi{10.3847/2041-8213/ac2c77}

\bibitem[{{Newton} {et~al.}(2017){Newton}, {Irwin}, {Charbonneau}, {Berlind},
  {Calkins}, \& {Mink}}]{Newton2017}
{Newton}, E.~R., {Irwin}, J., {Charbonneau}, D., {et~al.} 2017, \apj, 834, 85,
  \dodoi{10.3847/1538-4357/834/1/85}

\bibitem[{{Ning} {et~al.}(2018){Ning}, {Wolfgang}, \& {Ghosh}}]{Ning2018}
{Ning}, B., {Wolfgang}, A., \& {Ghosh}, S. 2018, \apj, 869, 5,
  \dodoi{10.3847/1538-4357/aaeb31}

\bibitem[{{Owen} \& {Campos Estrada}(2020)}]{Owen2020a}
{Owen}, J.~E., \& {Campos Estrada}, B. 2020, \mnras, 491, 5287,
  \dodoi{10.1093/mnras/stz3435}

\bibitem[{{Owen} {et~al.}(2020){Owen}, {Shaikhislamov}, {Lammer}, {Fossati}, \&
  {Khodachenko}}]{Owen2020b}
{Owen}, J.~E., {Shaikhislamov}, I.~F., {Lammer}, H., {Fossati}, L., \&
  {Khodachenko}, M.~L. 2020, \ssr, 216, 129, \dodoi{10.1007/s11214-020-00756-w}

\bibitem[{{Owen} \& {Wu}(2013)}]{Owen&Wu2013}
{Owen}, J.~E., \& {Wu}, Y. 2013, \apj, 775, 105,
  \dodoi{10.1088/0004-637X/775/2/105}

\bibitem[{Pidhorodetska {et~al.}(2021)Pidhorodetska, Moran, Schwieterman,
  Barclay, Fauchez, Lewis, Quintana, Villanueva, Domagal-Goldman, Schlieder,
  {et~al.}}]{Pidhorodetska2021}
Pidhorodetska, D., Moran, S.~E., Schwieterman, E.~W., {et~al.} 2021, \aj, 162,
  169, \dodoi{10.3847/1538-3881/ac1171}

\bibitem[{{Redfield} \& {Linsky}(2000)}]{Redfield2000}
{Redfield}, S., \& {Linsky}, J.~L. 2000, \apj, 534, 825, \dodoi{10.1086/308769}

\bibitem[{{Rogers}(2015)}]{Rogers2015}
{Rogers}, L.~A. 2015, \apj, 801, 41, \dodoi{10.1088/0004-637X/801/1/41}

\bibitem[{{Rogers} {et~al.}(2011){Rogers}, {Bodenheimer}, {Lissauer}, \&
  {Seager}}]{Rogers2011}
{Rogers}, L.~A., {Bodenheimer}, P., {Lissauer}, J.~J., \& {Seager}, S. 2011,
  \apj, 738, 59, \dodoi{10.1088/0004-637X/738/1/59}

\bibitem[{{Rogers} \& {Seager}(2010{\natexlab{a}})}]{Rogers&Seager2010a}
{Rogers}, L.~A., \& {Seager}, S. 2010{\natexlab{a}}, \apj, 712, 974

\bibitem[{{Rogers} \& {Seager}(2010{\natexlab{b}})}]{Rogers&Seager2010b}
---. 2010{\natexlab{b}}, \apj, 716, 1208, \dodoi{10.1088/0004-637X/716/2/1208}

\bibitem[{{Salz} {et~al.}(2016){Salz}, {Schneider}, {Czesla}, \&
  {Schmitt}}]{Salz2016}
{Salz}, M., {Schneider}, P.~C., {Czesla}, S., \& {Schmitt}, J.~H.~M.~M. 2016,
  \aap, 585, L2, \dodoi{10.1051/0004-6361/201527042}

\bibitem[{{Sanz-Forcada} {et~al.}(2003){Sanz-Forcada}, {Brickhouse}, \&
  {Dupree}}]{Sanz-Forcada2003}
{Sanz-Forcada}, J., {Brickhouse}, N.~S., \& {Dupree}, A.~K. 2003, \apjs, 145,
  147, \dodoi{10.1086/345815}

\bibitem[{{Schlichting} {et~al.}(2015){Schlichting}, {Sari}, \&
  {Yalinewich}}]{Schlichting2015}
{Schlichting}, H.~E., {Sari}, R., \& {Yalinewich}, A. 2015, \icarus, 247, 81,
  \dodoi{10.1016/j.icarus.2014.09.053}

\bibitem[{{Smith} {et~al.}(2001){Smith}, {Brickhouse}, {Liedahl}, \&
  {Raymond}}]{Smith2001}
{Smith}, R.~K., {Brickhouse}, N.~S., {Liedahl}, D.~A., \& {Raymond}, J.~C.
  2001, \apjl, 556, L91, \dodoi{10.1086/322992}

\bibitem[{{Speagle}(2020)}]{Speagle2020}
{Speagle}, J.~S. 2020, \mnras, 493, 3132, \dodoi{10.1093/mnras/staa278}

\bibitem[{{Tayar} {et~al.}(2022){Tayar}, {Claytor}, {Huber}, \& {van
  Saders}}]{Tayar2022}
{Tayar}, J., {Claytor}, Z.~R., {Huber}, D., \& {van Saders}, J. 2022, \apj,
  927, 31, \dodoi{10.3847/1538-4357/ac4bbc}

\bibitem[{Tsai {et~al.}(2017)Tsai, Lyons, Grosheintz, Rimmer, Kitzmann, \&
  Heng}]{Tsai2017}
Tsai, S.-M., Lyons, J.~R., Grosheintz, L., {et~al.} 2017, \apjs, 228, 20,
  \dodoi{10.3847/1538-4365/228/2/20}

\bibitem[{Tsai {et~al.}(2021)Tsai, Malik, Kitzmann, Lyons, Fateev, Lee, \&
  Heng}]{Tsai2021}
Tsai, S.-M., Malik, M., Kitzmann, D., {et~al.} 2021, \apj, 923, 264,
  \dodoi{10.3847/1538-4357/ac29bc}

\bibitem[{Turbet {et~al.}(2019)Turbet, Ehrenreich, Lovis, Bolmont, \&
  Fauchez}]{Turbet2019}
Turbet, M., Ehrenreich, D., Lovis, C., Bolmont, E., \& Fauchez, T. 2019, \aap,
  628, A12, \dodoi{10.1051/0004-6361/201935585}

\bibitem[{{Van Eylen} {et~al.}(2018){Van Eylen}, {Agentoft}, {Lundkvist},
  {Kjeldsen}, {Owen}, {Fulton}, {Petigura}, \& {Snellen}}]{VanEylen2018}
{Van Eylen}, V., {Agentoft}, C., {Lundkvist}, M.~S., {et~al.} 2018, \mnras,
  479, 4786, \dodoi{10.1093/mnras/sty1783}

\bibitem[{Villanueva {et~al.}(2018)Villanueva, Smith, Protopapa, Faggi, \&
  Mandell}]{Villanueva2018}
Villanueva, G.~L., Smith, M.~D., Protopapa, S., Faggi, S., \& Mandell, A.~M.
  2018, Journal of Quantitative Spectroscopy and Radiative Transfer, 217, 86,
  \dodoi{10.1016/j.jqsrt.2018.05.023}

\bibitem[{Way \& Del~Genio(2020)}]{Way&DelGenio2020}
Way, M.~J., \& Del~Genio, A.~D. 2020, Journal of Geophysical Research: Planets,
  125, e2019JE006276, \dodoi{10.1029/2019JE006276}

\bibitem[{{Wright} {et~al.}(2018){Wright}, {Newton}, {Williams}, {Drake}, \&
  {Yadav}}]{Wright2018}
{Wright}, N.~J., {Newton}, E.~R., {Williams}, P. K.~G., {Drake}, J.~J., \&
  {Yadav}, R.~K. 2018, \mnras, 479, 2351, \dodoi{10.1093/mnras/sty1670}

\bibitem[{{Wyatt} {et~al.}(2020){Wyatt}, {Kral}, \& {Sinclair}}]{Wyatt2020}
{Wyatt}, M.~C., {Kral}, Q., \& {Sinclair}, C.~A. 2020, \mnras, 491, 782,
  \dodoi{10.1093/mnras/stz3052}

\bibitem[{Yang {et~al.}(2019)Yang, Abbot, Koll, Hu, \& Showman}]{Yang2019}
Yang, J., Abbot, D.~S., Koll, D.~D., Hu, Y., \& Showman, A.~P. 2019, \apj, 871,
  29, \dodoi{10.3847/1538-4357/aaf1a8}

\bibitem[{Yang {et~al.}(2013)Yang, Cowan, \& Abbot}]{Yang2013}
Yang, J., Cowan, N.~B., \& Abbot, D.~S. 2013, \apjl, 771, L45,
  \dodoi{10.1088/2041-8205/771/2/L45}

\bibitem[{{Youngblood} {et~al.}(2016){Youngblood}, {France}, {Loyd}, {Linsky},
  {Redfield}, {Schneider}, {Wood}, {Brown}, {Froning}, {Miguel}, {Rugheimer},
  \& {Walkowicz}}]{Youngblood2016}
{Youngblood}, A., {France}, K., {Loyd}, R.~O.~P., {et~al.} 2016, \apj, 824,
  101, \dodoi{10.3847/0004-637X/824/2/101}

\bibitem[{{Youngblood} {et~al.}(2017){Youngblood}, {France}, {Loyd}, {Brown},
  {Mason}, {Schneider}, {Tilley}, {Berta-Thompson}, {Buccino}, {Froning},
  {Hawley}, {Linsky}, {Mauas}, {Redfield}, {Kowalski}, {Miguel}, {Newton},
  {Rugheimer}, {Segura}, {Roberge}, \& {Vieytes}}]{Youngblood2017}
---. 2017, \apj, 843, 31, \dodoi{10.3847/1538-4357/aa76dd}

\bibitem[{Zahnle {et~al.}(2008)Zahnle, Haberle, Catling, \&
  Kasting}]{Zahnle2008}
Zahnle, K., Haberle, R.~M., Catling, D.~C., \& Kasting, J.~F. 2008, JGR:P, 113,
  \dodoi{10.1029/2008JE003160}

\bibitem[{{Zeng} {et~al.}(2019){Zeng}, {Jacobsen}, {Sasselov}, {Petaev},
  {Vanderburg}, {Lopez-Morales}, {Perez-Mercader}, {Mattsson}, {Li}, {Heising},
  {Bonomo}, {Damasso}, {Berger}, {Cao}, {Levi}, \& {Wordsworth}}]{Zeng2019}
{Zeng}, L., {Jacobsen}, S.~B., {Sasselov}, D.~D., {et~al.} 2019, Proceedings of
  the National Academy of Science, 116, 9723, \dodoi{10.1073/pnas.1812905116}

\end{thebibliography}

\end{document}